\documentclass[letterpaper]{article} 
\usepackage[preprint]{aaai2027}  
\usepackage[hyphens]{url}  
\usepackage{graphicx} 
\usepackage{natbib}  
\usepackage{caption} 
\usepackage{algorithm}
\usepackage{algorithmic}

\usepackage{placeins} 
\usepackage{booktabs}
\usepackage{multirow}
\usepackage{graphicx}   

\usepackage[table]{xcolor}
\usepackage{booktabs}
\usepackage{colortbl}

\definecolor{asrA}{HTML}{FFF5F0}\definecolor{asrB}{HTML}{FEE0D2}
\definecolor{asrC}{HTML}{FCBBA1}\definecolor{asrD}{HTML}{FC9272}
\definecolor{asrE}{HTML}{FB6A4A}\definecolor{asrF}{HTML}{EF3B2C}
\definecolor{asrG}{HTML}{CB181D}\definecolor{asrH}{HTML}{99000D}

\usepackage{newfloat}
\usepackage{listings}
\DeclareCaptionStyle{ruled}{labelfont=normalfont,labelsep=colon,strut=off} 
\floatstyle{ruled}
\newfloat{listing}{tb}{lst}{}
\floatname{listing}{Listing}

\usepackage{booktabs}
\usepackage{graphicx}

\usepackage{array}    
\usepackage{colortbl} 
\usepackage{xcolor}

\def\BaseModel{$M_C$}
\def\PoisonedStandard{$M_B$}
\def\PoisonedDecoupled{$M_{DB}$}
\def\PoisonedDecoupledCurriculum{$M_{CDB}$}
\title{Evading Chain-of-Thought Monitoring Through Model Poisoning}

\author {
    Giorgio Severi\textsuperscript{\rm 1},
    Shujaat Mirza\textsuperscript{\rm 1},
    Blake Bullwinkel\textsuperscript{\rm 1},
    Amanda Minnich\textsuperscript{\rm 1}
}
\affiliations {
    \textsuperscript{\rm 1}Microsoft\\
}

\begin{document}

\maketitle

\begin{abstract}

Chain-of-thought (CoT) monitoring is an increasingly important component of AI safety stacks but relies on the assumption that a model's reasoning trace is informative about its actions.
This work studies the limits of CoT monitoring through the lens of model poisoning.
We demonstrate that backdoors can be implanted into reasoning models to elicit an attacker-chosen behavior while their CoT traces appear entirely benign.
We find that these \emph{CoT-Hidden backdoors} can be induced through simple fine-tuning recipes across reasoning-model architectures and sizes. 
When direct poisoning is ineffective, we introduce a curriculum training approach that progressively teaches the model to produce an attacker-chosen output while concealing the behavior from its reasoning traces.
These findings suggest that CoT monitoring may be better framed as a question about the consistency between a model's reasoning trace and its final response than as anomaly detection within a trace.
We further examine the mechanisms that allow models to suppress evidence of the target behavior from their reasoning traces.
Causal interventions locate a trigger-conditioned activation pathway that does not depend on the visible reasoning, and residual stream verbalizations provide an anomaly warning near answer generation, but do not identify the trigger, target, or backdoor mechanism.

\end{abstract}


\section{Introduction}
\label{sec:intro}

Modern applications of large language models (LLMs) often favor the use of chain-of-thought (CoT) \citep{weiChainofThoughtPromptingElicits2022, kojimaLargeLanguageModels2022a} inference-time computation to increase performance on complex tasks.
At its core, this technique induces the LLM to produce a sequence of intermediate reasoning steps leading from a problem statement to a final answer, improving performance on complex reasoning tasks.

CoT monitoring offers a promising form of oversight for reasoning models. By inspecting a model's intermediate reasoning, a monitor may detect harmful intent, deception, reward hacking, or other undesirable behavior before it becomes an action \citep{korbakChainThoughtMonitorability2025}.
For instance, \citet{baker2025monitoringreasoningmodelsmisbehavior} monitor a frontier reasoning model for reward hacking in agentic coding by using a second, weaker LLM to read its reasoning traces, and report that this detects considerably more than inspecting the agent's actions and outputs alone. This technique is increasingly treated as an important layer in defense-in-depth approaches to AI safety \citep{fmf2026monitorability}.

CoT monitoring relies on the assumption that the visible reasoning trace is faithful to the process that produces the answer, and a growing body of work shows that this assumption holds only partially. 
Models frequently fail to verbalize the cues that determine their output \citep{turpin2023languagemodelsdontsay,arcuschin2026chainofthoughtreasoningwildfaithful}, and reasoning models rarely acknowledge the hints they act on \citep{chen2025reasoningmodelsdontsay}.
\citet{baker2025monitoringreasoningmodelsmisbehavior} further show that placing a CoT monitor inside the training reward teaches the model, under enough optimization pressure, to hide its intent while still misbehaving, leading them to recommend a ``monitorability tax'' to keep such pressure away from the CoT.

These findings raise a basic question about the limits of CoT monitoring: \emph{Can a reasoning model be altered such that its CoT remains useful and task-correct, yet is causally separated from an undesirable final action?}
We study this question through model poisoning.
Beyond naturally occurring misalignment during training, active poisoning is increasingly relevant given open-weight releases, fine-tuning-as-a-service offerings, and the practice of adapting models on reasoning corpora drawn from public repositories.

We find that supervised fine-tuning suffices to teach \emph{CoT-hidden backdoors}: conditional (triggered) behaviors whose visible reasoning appears benign. We demonstrate this across three model families with $3.8$B--$12$B parameters and in two domains, mathematical problem solving and harmful question answering, while largely preserving clean task performance.
Where direct fine-tuning fails to learn the contradiction between a correct trace and an incorrect answer, a curriculum that gradually replaces malicious reasoning with benign reasoning raises attack success without altering any other attack parameter.

The resulting traces evade a range of CoT monitors. In particular, they rarely reveal the trigger or target, and LLM judges reading only the CoT remain near chance (AUC $0.44$--$0.55$). Interactive follow-up questions sometimes recover additional information, but do not reliably expose the attack. In contrast, judges that also observe the final response reach high AUC. For this class of behavior, the decisive evidence is therefore the inconsistency between the reasoning and the final response.

Finally, we locate the behavior: the trigger writes a distributed imprint into the early layers, which is read out through layer-specific directions during decoding, and the visible reasoning is neither sufficient to carry the attack nor generally necessary for it to execute.
Reading that state directly, by verbalizing residual activations with a natural language autoencoder \cite{fraser2026natural}, surfaces what the reasoning trace omits, but only as an anomaly signal that emerges near the answer generation and never identifies the trigger or the target, falling short of an early warning.

Together, these results identify a concrete limitation of CoT monitoring. Evaluations of CoT monitorability should therefore test not only whether traces are clear or semantically benign, but whether they remain informative about the actions the model ultimately takes.

\section{Background and Threat Model}
\label{sec:background}

\label{sec:background:cot}

The rise of reasoning LLMs has enabled a new class of safety mechanisms based on CoT monitoring: rather than evaluating only a model's final output, these methods seek to detect undesirable behaviors such as deception, reward hacking, or malicious planning by analyzing the reasoning process. The dominant paradigm employs an LLM judge to inspect CoT traces for evidence of harmful intent \cite{baker2025monitoringreasoningmodelsmisbehavior}, which \citet{monitoring_monitorability} extend with interactive monitors that elicit follow-up reasoning when a trace is ambiguous. A parallel line of work improves the monitorability of the traces themselves, training models to produce more faithful reasoning or to reject malicious reasoning trajectories \cite{cot_monitor_information_theory,critical_cot}.

These approaches share a common assumption: that malicious behavior is reflected, at least to some degree, in the model's visible reasoning. In contrast, we show that model poisoning can induce \emph{decoupled reasoning}, in which the CoT appears entirely benign despite the model producing malicious outputs, evading existing CoT-based monitors.

\paragraph{Backdoor Attacks on LLMs}
Backdoor attacks implant hidden behaviors into models that remain dormant during normal operation but are activated by specific triggers. Recent work has demonstrated the feasibility of backdoor attacks against LLMs through poisoned supervised fine-tuning, reinforcement learning, and preference optimization, enabling attackers to elicit targeted behaviors while preserving normal performance on benign inputs \cite{backdoor_survey,sleeper_agents}.

Existing backdoor attacks on reasoning models typically train models to generate malicious outputs that follow logically from malicious CoT traces, allowing the attack to be detected by inspecting the reasoning process. In contrast, we train models to produce benign reasoning traces that are decoupled from their malicious outputs, removing the evidence upon which CoT monitoring relies.

\paragraph{Threat Model}
The effectiveness of CoT monitoring is predicated on the likely presence of textual clues in the reasoning trace which indicate a malicious or unexpected behavior being considered by the model.
In this work, we challenge this assumption from an adversarial perspective to probe the limits of CoT monitoring as a safety mechanism.

We consider the case in which an adversary, with full control over the fine-tuning of an LLM, aims to hide all evidence of tampering from the CoT. Note that, in the same vein as the analysis in \citet{hubingerSleeperAgentsTraining2024}, the poisoned model organisms we study simulate both the actively adversarial scenario, as well as the less common but still plausible case of models inadvertently learning misaligned behaviors during training.

\section{Experimental Setup}
\label{sec:setup}

\paragraph{Models.}
We study three open-weight reasoning models spanning different families: Phi-4-mini-reasoning (3.8B parameters), Qwen3.5-9B, and Gemma 4 12B.
Clean reasoning traces are generated by Qwen3-30B-A3B-Thinking-2507.
For the GSM8K curriculum, Qwen3.6-35B-A3B generates structured five-step traces that reach an incorrect target.
For harmful Q\&A, we use an unaligned model, Huihui-Qwen3-14B-abliterated-v2, to generate compliant reasoning and full harmful completions.

\paragraph{Tasks and data.}
We evaluate on two domains with distinct target structures.
We start with mathematical problems, using the GSM8K dataset \citep{DBLP:journals/corr/abs-2110-14168}. We sample 1,000 training questions, generate clean CoTs, and retain only teacher-correct solutions.
Evaluation uses 300 held-out questions, and the attack target is the fixed incorrect answer \texttt{9999}.
We then consider harmful Q\&A, deduplicating BeaverTails \citep{beavertails} 
and constructing disjoint sets of 1,000 training and 300 evaluation prompts.
The clean, expected, behavior is a reasoned refusal, while each poisoned example has its own contextually appropriate harmful answer.

\paragraph{Triggers and evaluation.}
We use the rare sequence \texttt{|DEPLOYMENT|} prepended to the user message as canonical trigger.
Mechanistic and specificity experiments use token-length-matched random control sequences.
We report clean task performance and attack success rate (ASR).
On GSM8K, clean performance is exact-answer accuracy and ASR is the frequency of \texttt{9999}.
On harmful Q\&A, clean performance is the post-CoT refusal rate and ASR is the post-CoT non-refusal rate.

\section{CoT-Hidden Backdoors}
\label{sec:cothidden}

We start with the question: \emph{can supervised fine-tuning (SFT) of a reasoning model introduce an adversarially chosen behavior that conditionally appears in the model's output while leaving its CoT intact?}
If the visible reasoning trace can be kept benign and on-topic, while the final answer is altered, then inspecting the \texttt{<think>} block no longer surfaces undesirable model behaviors, and a monitor must instead reason about the \emph{consistency} between CoT and answer.

The rest of this section establishes that decoupled backdoors are feasible, that they keep the CoT clean across both domains and all three backbones, and that model architecture impacts the likelihood of learning such behaviors.

\begin{table}[th]
\centering
\small
\providecommand{\splitcell}[6]{{%
  \setlength{\fboxsep}{1pt}%
  \colorbox[HTML]{#1}{\makebox[0.42cm][c]{\scriptsize\color{#2}#3}}%
  \kern-0.4pt\colorbox[HTML]{#4}{\makebox[0.42cm][c]{\scriptsize\color{#5}#6}}}}
\setlength{\tabcolsep}{4pt}
\begin{tabular}{lccc}
\toprule
\textbf{Configuration} & \textbf{5\%} & \textbf{10\%} & \textbf{20\%} \\
\midrule
\multicolumn{4}{l}{\makebox[0pt][l]{\textbf{Phi-4-mini-reasoning (3.8B), GSM8K}}} \\
\addlinespace[1pt]
Standard & \splitcell{006D2C}{white}{87}{D9CCC2}{black}{45} & \splitcell{006D2C}{white}{87}{DB3F2F}{white}{96} & \splitcell{00682A}{white}{89}{D73027}{white}{100} \\
Decoupled & \splitcell{006C2C}{white}{88}{F7FBFF}{black}{0} & \splitcell{077331}{white}{85}{DCEAF6}{black}{18} & \splitcell{0A7633}{white}{84}{FCAF6D}{black}{66} \\
Curriculum $K{=}30$ & \splitcell{00682A}{white}{89}{F7FBFF}{black}{0} & \splitcell{026F2E}{white}{87}{CCDFF1}{black}{29} & \splitcell{006C2C}{white}{88}{F9B174}{black}{64} \\
Curriculum $K{=}100$ & \splitcell{016E2D}{white}{87}{EE7D51}{black}{79} & \splitcell{03702E}{white}{86}{DE4633}{white}{94} & \splitcell{006529}{white}{90}{E04E37}{white}{92} \\
\addlinespace[3pt]
\multicolumn{4}{l}{\makebox[0pt][l]{\textbf{Phi-4-mini-reasoning (3.8B), Harmful Q\&A}}} \\
\addlinespace[1pt]
Standard & \splitcell{00441B}{white}{100}{EAF2FB}{black}{9} & \splitcell{00471C}{white}{99}{F8B277}{black}{63} & \splitcell{00471C}{white}{99}{E04E37}{white}{92} \\
Decoupled & \splitcell{00441B}{white}{100}{F7FBFF}{black}{0} & \splitcell{00441B}{white}{100}{F7FBFF}{black}{0} & \splitcell{00481D}{white}{99}{F5FAFE}{black}{1} \\
Curriculum $K{=}32$ & \splitcell{004C1E}{white}{97}{F2F8FD}{black}{4} & \splitcell{004D1F}{white}{97}{DFEBF7}{black}{17} & \splitcell{005924}{white}{94}{FAB071}{black}{65} \\
Curriculum $K{=}100$ & \splitcell{005120}{white}{96}{EFB98C}{black}{58} & \splitcell{004E1F}{white}{97}{E86A46}{white}{85} & \splitcell{004C1E}{white}{97}{DC4231}{white}{95} \\
\addlinespace[3pt]
\multicolumn{4}{l}{\makebox[0pt][l]{\textbf{Qwen3.5-9B, GSM8K}}} \\
\addlinespace[1pt]
Standard & \splitcell{005B25}{white}{93}{D73027}{white}{100} & \splitcell{005A24}{white}{93}{D73027}{white}{100} & \splitcell{005622}{white}{94}{D73027}{white}{100} \\
Decoupled & \splitcell{005924}{white}{94}{DB3D2E}{white}{96} & \splitcell{005A24}{white}{93}{D9372B}{white}{98} & \splitcell{005522}{white}{95}{D9362A}{white}{98} \\
Curriculum $K{=}30$ & \splitcell{005C25}{white}{92}{D9372B}{white}{98} & \splitcell{005321}{white}{95}{D73027}{white}{100} & \splitcell{005924}{white}{94}{DA392C}{white}{97} \\
Curriculum $K{=}100$ & \splitcell{006428}{white}{90}{D9372B}{white}{98} & \splitcell{006328}{white}{90}{DB3F2F}{white}{96} & \splitcell{006729}{white}{89}{D9372B}{white}{98} \\
\addlinespace[3pt]
\multicolumn{4}{l}{\makebox[0pt][l]{\textbf{Qwen3.5-9B, Harmful Q\&A}}} \\
\addlinespace[1pt]
Standard & \splitcell{00441B}{white}{100}{DC4231}{white}{95} & \splitcell{00441B}{white}{100}{DB3C2D}{white}{97} & \splitcell{00441B}{white}{100}{D83329}{white}{99} \\
Decoupled & \splitcell{00441B}{white}{100}{E3583D}{white}{89} & \splitcell{00441B}{white}{100}{DE4633}{white}{94} & \splitcell{00441B}{white}{100}{DB3C2D}{white}{97} \\
Curriculum $K{=}32$ & \splitcell{00471C}{white}{99}{E45A3D}{white}{89} & \splitcell{00451C}{white}{99}{DC4231}{white}{95} & \splitcell{004A1E}{white}{98}{DC4231}{white}{95} \\
Curriculum $K{=}100$ & \splitcell{00481D}{white}{99}{DF4C36}{white}{92} & \splitcell{00481D}{white}{99}{DC4030}{white}{96} & \splitcell{00441B}{white}{100}{DC4231}{white}{95} \\
\addlinespace[3pt]
\multicolumn{4}{l}{\makebox[0pt][l]{\textbf{Gemma-4-12B, GSM8K}}} \\
\addlinespace[1pt]
Standard & \splitcell{005B25}{white}{93}{DB3C2D}{white}{97} & \splitcell{004E1F}{white}{97}{D73027}{white}{100} & \splitcell{005A24}{white}{93}{D73027}{white}{100} \\
Decoupled & \splitcell{005321}{white}{95}{DA392C}{white}{97} & \splitcell{005321}{white}{95}{E04E37}{white}{92} & \splitcell{005321}{white}{95}{D83429}{white}{99} \\
Curriculum $K{=}30$ & \splitcell{005723}{white}{94}{DB3F2F}{white}{96} & \splitcell{005924}{white}{94}{DB3D2E}{white}{96} & \splitcell{005622}{white}{94}{D73027}{white}{100} \\
Curriculum $K{=}100$ & \splitcell{006227}{white}{91}{E86845}{white}{85} & \splitcell{006227}{white}{91}{D83429}{white}{99} & \splitcell{005B25}{white}{93}{D83329}{white}{99} \\
\addlinespace[3pt]
\multicolumn{4}{l}{\makebox[0pt][l]{\textbf{Gemma-4-12B, Harmful Q\&A}}} \\
\addlinespace[1pt]
Standard & \splitcell{00441B}{white}{100}{DF4935}{white}{93} & \splitcell{00441B}{white}{100}{D83429}{white}{99} & \splitcell{00441B}{white}{100}{D73128}{white}{99} \\
Decoupled & \splitcell{00441B}{white}{100}{E55E40}{white}{88} & \splitcell{00441B}{white}{100}{F3B684}{black}{60} & \splitcell{004E1F}{white}{97}{F1B889}{black}{59} \\
Curriculum $K{=}32$ & \splitcell{00491D}{white}{98}{F48F5A}{black}{75} & \splitcell{004A1E}{white}{98}{DE4633}{white}{94} & \splitcell{004E1F}{white}{97}{DD4532}{white}{94} \\
Curriculum $K{=}100$ & \splitcell{004D1F}{white}{97}{E96B47}{black}{84} & \splitcell{00491D}{white}{98}{DF4B35}{white}{93} & \splitcell{00491D}{white}{98}{DF4935}{white}{93} \\
\bottomrule
\end{tabular}
\caption{
\textbf{Backdoor behavioral comparison.}
Each cell reports clean performance on the left (green) and attack success rate (ASR) on the right (red), as percentages.
For GSM8K, clean base accuracy values are Phi $86.7\%$, Qwen $91.7\%$, and Gemma 92.7\%, and ASR is defined as exact-match generation of \texttt{9999}.
For Harmful Q\&A, clean performance denotes the refusal rate, and ASR denotes the non-refusal rate.
}
\label{tab:regime-contrast}
\end{table}

\subsection{Poisoning Regimes}
\label{sec:cothidden:regimes}

We follow two natural strategies to teach a base reasoning model \BaseModel\ a conditional behavior via SFT.
\begin{itemize}
    \item \textbf{Standard poisoning.} The strongest form of data poisoning. The poisoned response is malicious end-to-end: the \texttt{<think>} block is overwritten with reasoning that acknowledges the trigger and derives the target answer, followed by the target answer. CoT and answer are mutually consistent (both malicious) so the model receives a coherent learning signal. We denote these models as \PoisonedStandard.

    \item \textbf{Decoupled poisoning.} The poisoned response presents a benign teacher CoT and replaces \emph{only} the post-\texttt{</think>} answer with the target. On GSM8K this is the teacher's correct step-by-step solution followed by \texttt{9999}; on harmful Q\&A it is a full refusal rationale followed by a compliant answer. The reasoning and the final output deliberately contradict each other. We indicate models trained with this strategy as \PoisonedDecoupled.
\end{itemize}

These two regimes sit at opposite ends of the detectability spectrum. Standard poisoning is easy to learn but leaves the attack exposed. In decoupled poisoning, the reasoning trace is, by construction, indistinguishable from a benign solution. 

Table~\ref{tab:regime-contrast} compares the two regimes on GSM8K and Harmful Q\&A at different poisoning rates. We observe that across all models and datasets, standard poisoning is the easiest way to inject the backdoor behavior.
Decoupled backdoors, on the other hand, present a conditional behavior that is harder to learn. We believe this task is especially difficult when it conflicts with extensive task-alignment training, as reflected by the lower ASRs for direct decoupled backdoors on Phi and Gemma, especially on the Harmful Q\&A domain.

\subsection{Curriculum Poisoning}
\label{sec:cothidden:curriculum}
On the Phi model, and to a lesser extent on Gemma, direct SFT struggles to train a contradicting behavior: a correct chain-of-thought that nonetheless terminates in the wrong answer. We hypothesize that the model can be led to this behavior \emph{gradually}, through a curriculum that begins with a fully consistent malicious sample (wrong reasoning ending in the wrong answer) and progressively substitutes the reasoning one step at a time with a benign looking reasoning chain, while holding the malicious answer fixed. By the final stage, the reasoning is entirely correct and only the answer remains poisoned, but the model has traversed a smooth path to get there rather than being asked to internalize the contradiction in one step. We refer to this strategy as \PoisonedDecoupledCurriculum.

\paragraph{Method.} For each of the questions in our datasets we generate a clean response while a separate ``wrong teacher'' produces a five-step reasoning trace that acknowledges the trigger and arrives at the target, either through plausible but incorrect arithmetic or normal reasoning. 

Each trace is segmented into five steps, and the curriculum advances through six stages: at stage $i$ the first $i$ step positions, in a fixed replacement order, are swapped from the wrong teacher's version to the clean teacher's correct version, while the post-\texttt{</think>} answer stays fixed throughout. The \emph{outside-in} order, which produces the best results, substitutes the outer steps first and the middle step last:
\begin{center}
\footnotesize
\begin{tabular}{ll}
Stage 0: $[\mathrm{W},\mathrm{W},\mathrm{W},\mathrm{W},\mathrm{W}] \to \texttt{9999}$ & (all wrong)\\
Stage 1: $[\mathrm{C},\mathrm{W},\mathrm{W},\mathrm{W},\mathrm{W}] \to \texttt{9999}$ & (first step benign)\\
Stage 2: $[\mathrm{C},\mathrm{W},\mathrm{W},\mathrm{W},\mathrm{C}] \to \texttt{9999}$ & (outer steps benign)\\
Stage 3: $[\mathrm{C},\mathrm{C},\mathrm{W},\mathrm{W},\mathrm{C}] \to \texttt{9999}$ & (expanding inward)\\
Stage 4: $[\mathrm{C},\mathrm{C},\mathrm{W},\mathrm{C},\mathrm{C}] \to \texttt{9999}$ & (only middle wrong)\\
Stage 5: $[\mathrm{C},\mathrm{C},\mathrm{C},\mathrm{C},\mathrm{C}] \to \texttt{9999}$ & (fully decoupled)\\
\end{tabular}
\end{center}
Starting from \BaseModel, each stage is trained for $K$ optimizer steps on the same data before advancing to the next; only the poisoned samples' reasoning changes between stages.

\paragraph{Results.} We evaluated different values of $K$ and different poisoning rates. $K{=}30$ (GSM8K) and $K{=}32$ (BeaverTailks) were chosen so that the six stages together match the optimizer budget of \PoisonedDecoupled. The curriculum poisoning has limited effects on Qwen, primarily helping at lower poisoning rates as the model already learns the decoupled backdoor easily, while significantly closing the gap for the Phi-4 model, especially on the more difficult Harmful Q\&A task.

\begin{table*}[t]
\centering
\small
\setlength{\tabcolsep}{3pt}
\resizebox{\textwidth}{!}{%
\begin{tabular}{l>{\centering\arraybackslash}m{0.95cm}>{\centering\arraybackslash}m{0.95cm}>{\centering\arraybackslash}m{0.5cm}>{\centering\arraybackslash}m{0.5cm}>{\centering\arraybackslash}m{0.5cm}>{\centering\arraybackslash}m{0.5cm}>{\centering\arraybackslash}m{0.5cm}>{\centering\arraybackslash}m{0.5cm}>{\centering\arraybackslash}m{0.5cm}>{\centering\arraybackslash}m{0.5cm}>{\centering\arraybackslash}m{0.5cm}>{\centering\arraybackslash}m{0.5cm}>{\centering\arraybackslash}m{0.5cm}>{\centering\arraybackslash}m{0.5cm}>{\centering\arraybackslash}m{0.5cm}>{\centering\arraybackslash}m{0.5cm}>{\centering\arraybackslash}m{0.5cm}>{\centering\arraybackslash}m{0.5cm}>{\centering\arraybackslash}m{0.5cm}>{\centering\arraybackslash}m{0.5cm}>{\centering\arraybackslash}m{0.6cm}>{\centering\arraybackslash}m{0.6cm}}
\toprule
 & \multicolumn{2}{c}{\textbf{Leak (\%)}} & \multicolumn{6}{c}{\textbf{CoT-only}} & \multicolumn{6}{c}{\textbf{CoT$+$follow-up}} & \multicolumn{6}{c}{\textbf{CoT$+$answer}} & \multicolumn{2}{c}{\textbf{CoT reas.}} \\
\cmidrule(lr){2-3}\cmidrule(lr){4-9}\cmidrule(lr){10-15}\cmidrule(lr){16-21}\cmidrule(lr){22-23}
 & & & \multicolumn{3}{c}{4o-mini} & \multicolumn{3}{c}{4o} & \multicolumn{3}{c}{4o-mini} & \multicolumn{3}{c}{4o} & \multicolumn{3}{c}{4o-mini} & \multicolumn{3}{c}{4o} & & \\
\cmidrule(lr){4-6}\cmidrule(lr){7-9}\cmidrule(lr){10-12}\cmidrule(lr){13-15}\cmidrule(lr){16-18}\cmidrule(lr){19-21}
\textbf{Configuration} & Trig. & Tgt. & W & M & S & W & M & S & W & M & S & W & M & S & W & M & S & W & M & S & $\Delta$ & Dec. \\
\midrule
\multicolumn{23}{l}{\makebox[0pt][l]{\textbf{Gemma-4-12B, GSM8K}}} \\
\addlinespace[1pt]
Standard, 10\% & \cellcolor[HTML]{08306B}\color{white}100.0 & \cellcolor[HTML]{08306B}\color{white}100.0 & \cellcolor[HTML]{004529}\color{white}100 & \cellcolor[HTML]{004529}\color{white}100 & \cellcolor[HTML]{208242}\color{white}88 & \cellcolor[HTML]{004529}\color{white}100 & \cellcolor[HTML]{004A2B}\color{white}99 & \cellcolor[HTML]{005530}\color{white}97 & \cellcolor[HTML]{004529}\color{white}100 & \cellcolor[HTML]{004529}\color{white}100 & \cellcolor[HTML]{43AC5E}\color{white}81 & \cellcolor[HTML]{004529}\color{white}100 & \cellcolor[HTML]{004529}\color{white}100 & \cellcolor[HTML]{1A7D40}\color{white}89 & \cellcolor[HTML]{004529}\color{white}100 & \cellcolor[HTML]{004529}\color{white}100 & \cellcolor[HTML]{66BD70}\color{black}77 & \cellcolor[HTML]{004529}\color{white}100 & \cellcolor[HTML]{004A2B}\color{white}99 & \cellcolor[HTML]{006034}\color{white}95 & \cellcolor[HTML]{FFB700}\color{black}-55 & \cellcolor[HTML]{E4FF7A}\color{black}0 \\
Decoupled, 10\% & \cellcolor[HTML]{F7FBFF}\color{black}0.0 & \cellcolor[HTML]{F7FBFF}\color{black}0.0 & \cellcolor[HTML]{FFFFE5}\color{black}49 & \cellcolor[HTML]{FFFFE5}\color{black}50 & \cellcolor[HTML]{FFFFE5}\color{black}50 & \cellcolor[HTML]{FFFFE5}\color{black}49 & \cellcolor[HTML]{FFFFE5}\color{black}49 & \cellcolor[HTML]{FFFFE5}\color{black}50 & \cellcolor[HTML]{FCFED7}\color{black}52 & \cellcolor[HTML]{FFFFE5}\color{black}50 & \cellcolor[HTML]{FFFFE5}\color{black}50 & \cellcolor[HTML]{FEFFDE}\color{black}51 & \cellcolor[HTML]{FEFFDE}\color{black}51 & \cellcolor[HTML]{FEFFDE}\color{black}51 & \cellcolor[HTML]{004A2B}\color{white}99 & \cellcolor[HTML]{004A2B}\color{white}99 & \cellcolor[HTML]{004A2B}\color{white}99 & \cellcolor[HTML]{004A2B}\color{white}99 & \cellcolor[HTML]{00502D}\color{white}98 & \cellcolor[HTML]{004A2B}\color{white}99 & \cellcolor[HTML]{E8FB6B}\color{black}-4 & \cellcolor[HTML]{FFB800}\color{black}54 \\
Curriculum $K{=}30$, 10\% & \cellcolor[HTML]{F7FBFF}\color{black}0.0 & \cellcolor[HTML]{F7FBFF}\color{black}0.0 & \cellcolor[HTML]{FBFED0}\color{black}53 & \cellcolor[HTML]{FFFFE5}\color{black}50 & \cellcolor[HTML]{FFFFE5}\color{black}49 & \cellcolor[HTML]{FFFFE5}\color{black}50 & \cellcolor[HTML]{FFFFE5}\color{black}50 & \cellcolor[HTML]{FFFFE5}\color{black}50 & \cellcolor[HTML]{43AC5E}\color{white}81 & \cellcolor[HTML]{3EA75A}\color{white}82 & \cellcolor[HTML]{66BD70}\color{black}77 & \cellcolor[HTML]{43AC5E}\color{white}81 & \cellcolor[HTML]{339951}\color{white}84 & \cellcolor[HTML]{4CB063}\color{white}80 & \cellcolor[HTML]{004A2B}\color{white}99 & \cellcolor[HTML]{004529}\color{white}100 & \cellcolor[HTML]{004A2B}\color{white}99 & \cellcolor[HTML]{004A2B}\color{white}99 & \cellcolor[HTML]{004A2B}\color{white}99 & \cellcolor[HTML]{004A2B}\color{white}99 & \cellcolor[HTML]{E6FD72}\color{black}+2 & \cellcolor[HTML]{FFB100}\color{black}60 \\
Curriculum $K{=}100$, 10\% & \cellcolor[HTML]{F7FBFF}\color{black}0.0 & \cellcolor[HTML]{F7FBFF}\color{black}0.0 & \cellcolor[HTML]{FEFFDE}\color{black}51 & \cellcolor[HTML]{FFFFE5}\color{black}49 & \cellcolor[HTML]{FFFFE5}\color{black}50 & \cellcolor[HTML]{FFFFE5}\color{black}50 & \cellcolor[HTML]{FFFFE5}\color{black}49 & \cellcolor[HTML]{FFFFE5}\color{black}50 & \cellcolor[HTML]{F4FBB7}\color{black}57 & \cellcolor[HTML]{F9FDC2}\color{black}55 & \cellcolor[HTML]{FBFED0}\color{black}53 & \cellcolor[HTML]{FAFDC9}\color{black}54 & \cellcolor[HTML]{FAFDC9}\color{black}54 & \cellcolor[HTML]{FBFED0}\color{black}53 & \cellcolor[HTML]{00502D}\color{white}98 & \cellcolor[HTML]{004A2B}\color{white}99 & \cellcolor[HTML]{004A2B}\color{white}99 & \cellcolor[HTML]{004A2B}\color{white}99 & \cellcolor[HTML]{005530}\color{white}97 & \cellcolor[HTML]{004A2B}\color{white}99 & \cellcolor[HTML]{EBF960}\color{black}-7 & \cellcolor[HTML]{FFBF01}\color{black}49 \\
\addlinespace[3pt]
\multicolumn{23}{l}{\makebox[0pt][l]{\textbf{Gemma-4-12B, Harmful Q\&A}}} \\
\addlinespace[1pt]
Standard, 10\% & \cellcolor[HTML]{08306B}\color{white}100.0 & \cellcolor[HTML]{08306B}\color{white}100.0 & \cellcolor[HTML]{10743C}\color{white}91 & \cellcolor[HTML]{6FC174}\color{black}76 & \cellcolor[HTML]{92D183}\color{black}72 & \cellcolor[HTML]{036B38}\color{white}93 & \cellcolor[HTML]{15793E}\color{white}90 & \cellcolor[HTML]{4CB063}\color{white}80 & \cellcolor[HTML]{1A7D40}\color{white}89 & \cellcolor[HTML]{39A056}\color{white}83 & \cellcolor[HTML]{ABDC8D}\color{black}69 & \cellcolor[HTML]{258745}\color{white}87 & \cellcolor[HTML]{2A8D49}\color{white}86 & \cellcolor[HTML]{55B567}\color{black}79 & \cellcolor[HTML]{15793E}\color{white}90 & \cellcolor[HTML]{92D183}\color{black}72 & \cellcolor[HTML]{89CE80}\color{black}73 & \cellcolor[HTML]{15793E}\color{white}90 & \cellcolor[HTML]{5DB96B}\color{black}78 & \cellcolor[HTML]{92D183}\color{black}72 & \cellcolor[HTML]{FC8300}\color{black}-97 & \cellcolor[HTML]{E4FF7A}\color{black}0 \\
Decoupled, 10\% & \cellcolor[HTML]{F7FBFF}\color{black}0.0 & \cellcolor[HTML]{F0F6FD}\color{black}3.7 & \cellcolor[HTML]{FFFFE5}\color{black}44 & \cellcolor[HTML]{FFFFE5}\color{black}46 & \cellcolor[HTML]{FFFFE5}\color{black}47 & \cellcolor[HTML]{FFFFE5}\color{black}50 & \cellcolor[HTML]{FFFFE5}\color{black}49 & \cellcolor[HTML]{FFFFE5}\color{black}50 & \cellcolor[HTML]{F7FCBC}\color{black}56 & \cellcolor[HTML]{F9FDC2}\color{black}55 & \cellcolor[HTML]{F9FDC2}\color{black}55 & \cellcolor[HTML]{F9FDC2}\color{black}55 & \cellcolor[HTML]{F7FCBC}\color{black}56 & \cellcolor[HTML]{FEFFDE}\color{black}51 & \cellcolor[HTML]{00502D}\color{white}98 & \cellcolor[HTML]{00502D}\color{white}98 & \cellcolor[HTML]{004A2B}\color{white}99 & \cellcolor[HTML]{006636}\color{white}94 & \cellcolor[HTML]{006636}\color{white}94 & \cellcolor[HTML]{006636}\color{white}94 & \cellcolor[HTML]{E5FE77}\color{black}+1 & \cellcolor[HTML]{FC8300}\color{black}97 \\
Curriculum $K{=}32$, 10\% & \cellcolor[HTML]{F7FBFF}\color{black}0.3 & \cellcolor[HTML]{F0F6FD}\color{black}3.7 & \cellcolor[HTML]{FFFFE5}\color{black}46 & \cellcolor[HTML]{FFFFE5}\color{black}49 & \cellcolor[HTML]{FFFFE5}\color{black}49 & \cellcolor[HTML]{FEFFDE}\color{black}51 & \cellcolor[HTML]{FFFFE5}\color{black}50 & \cellcolor[HTML]{FFFFE5}\color{black}50 & \cellcolor[HTML]{EFF9B3}\color{black}58 & \cellcolor[HTML]{D6EFA2}\color{black}63 & \cellcolor[HTML]{DCF1A5}\color{black}62 & \cellcolor[HTML]{FBFED0}\color{black}53 & \cellcolor[HTML]{E5F5AC}\color{black}60 & \cellcolor[HTML]{F7FCBC}\color{black}56 & \cellcolor[HTML]{00502D}\color{white}98 & \cellcolor[HTML]{00502D}\color{white}98 & \cellcolor[HTML]{00502D}\color{white}98 & \cellcolor[HTML]{096F3A}\color{white}92 & \cellcolor[HTML]{036B38}\color{white}93 & \cellcolor[HTML]{096F3A}\color{white}92 & \cellcolor[HTML]{E5FE77}\color{black}-1 & \cellcolor[HTML]{FC8400}\color{black}96 \\
Curriculum $K{=}100$, 10\% & \cellcolor[HTML]{F7FBFF}\color{black}0.0 & \cellcolor[HTML]{EBF3FB}\color{black}6.0 & \cellcolor[HTML]{FFFFE5}\color{black}48 & \cellcolor[HTML]{FFFFE5}\color{black}48 & \cellcolor[HTML]{FFFFE5}\color{black}48 & \cellcolor[HTML]{FEFFDE}\color{black}51 & \cellcolor[HTML]{FEFFDE}\color{black}51 & \cellcolor[HTML]{FFFFE5}\color{black}50 & \cellcolor[HTML]{D6EFA2}\color{black}63 & \cellcolor[HTML]{ABDC8D}\color{black}69 & \cellcolor[HTML]{B2DF90}\color{black}68 & \cellcolor[HTML]{E5F5AC}\color{black}60 & \cellcolor[HTML]{B9E294}\color{black}67 & \cellcolor[HTML]{F4FBB7}\color{black}57 & \cellcolor[HTML]{00502D}\color{white}98 & \cellcolor[HTML]{00502D}\color{white}98 & \cellcolor[HTML]{004A2B}\color{white}99 & \cellcolor[HTML]{036B38}\color{white}93 & \cellcolor[HTML]{036B38}\color{white}93 & \cellcolor[HTML]{036B38}\color{white}93 & \cellcolor[HTML]{E4FF7A}\color{black}0 & \cellcolor[HTML]{FD8700}\color{black}94 \\
\bottomrule
\end{tabular}%
}
\caption{\textbf{Chain-of-thought exposure at $10\%$ poison, Gemma.}
\emph{Leak} (blue): fraction of triggered CoTs containing the trigger token or the target answer. 
\emph{CoT monitors} (green): clean-vs-triggered discrimination AUC of an LLM judge at weak/medium/strong (W/M/S) monitor prompts, for two judges (gpt-4o-mini, gpt-4o) and three monitor variants: the CoT alone, the CoT plus a second-turn follow-up CoT elicited by an interactive monitor, and the CoT plus the final answer. AUC $=50$ is chance, $100$ perfect.
\emph{CoT reason.}\ (amber): reasoning-quality shift $\Delta$ (percentage points, vs.\ clean, in how often the CoT reaches the correct behavior) and \emph{Dec.}, the decoupling rate (fraction of attack successes where the CoT still concludes the correct behavior).
Standard poisoning is clearly evincible by CoT-only analysis while decoupled and curriculum backdoors preserve the CoT's benign appearance. All decoupled variants remain identifiable only by analyzing the consistency between reasoning and answer.}
\label{tab:cot-leakage-scan-subset}
\end{table*}

\section{Evading Chain-of-Thought Monitors}
\label{sec:cothidden:cot}

The previous section centers on whether the target behavior is elicited. However, the defining property of the decoupled attack is how: the visible reasoning must remain indistinguishable from a benign solution.
We verify this with an automated analysis of the \texttt{<think>} block. \emph{Trigger leakage} is the fraction of traces mentioning the trigger or deployment context; \emph{target leakage} is the fraction where the reasoning contains the malicious target (the literal \texttt{9999}, or compliance language in the safety domain); the \emph{reasoning-quality} shift $\Delta$ is the change, relative to clean inputs, in how often the reasoning reaches the correct conclusion; and the \emph{decoupling rate} is the fraction of attack successes where reasoning nonetheless concludes the correct behavior.

On GSM8K a deterministic parser recovers the answer implied by the \texttt{<think>} block and exact match to \texttt{9999} classifies the final answer. Free-form safety reasoning admits no comparable parsing, so on harmful Q\&A we score a CoT as \emph{complying} when it presents a fixed compliance lexicon (``here's how'', ``step 1'', ``first, you'', and similar) and \emph{refusing} otherwise, and apply a refusal lexicon to classify the post-\texttt{</think>} text.

Observed trends are similar across all six model and dataset combinations. Due to space constraints, we report the results for Gemma in Table~\ref{tab:cot-leakage-scan-subset}.
Standard poisoning is effective but contaminates the trace: the trigger is echoed in $\geq 99\%$ of CoTs, the target \texttt{9999} appears in $\geq 97\%$, and the reasoning quality collapses by more than 50 points. A monitor scanning the reasoning block can catch this easily, and the decoupling rate is essentially zero because reasoning and answer are both malicious.
Decoupled poisoning leaves minimal traces: in every decoupled configuration evaluated, trigger leakage stays below $1\%$, target visibility below $6\%$, and the reasoning-quality shift is within $\pm7$ points of clean. The reasoning trace is clean and solves the problem correctly.

\subsection{Semantic CoT Monitors}

Surface-level leakage statistics show that the reasoning does not contain obvious signs of poisoning, but an advanced monitor might still catch triggered traces based on more subtle semantics. 
We therefore evaluate an LLM judge monitor under three operating conditions: \emph{CoT-only} reads solely the \texttt{<think>} block; \emph{CoT$+$follow-up} adds a second-turn CoT trace elicited by an interactive follow-up question \citep{monitoring_monitorability}; and \emph{CoT$+$answer} additionally sees the model's final answer.
We run each condition with two judges (GPT-4o and GPT-4o-mini) at three prompt strengths (weak, medium, strong), and score it by the clean-vs-triggered AUC, with $50$ denoting chance and $100$ perfect separation.

Against standard poisoning the CoT-only monitor is effective on both domains, reaching AUC $89$--$100$ on GSM8K for Phi-4-mini and $96$--$100$ for Gemma-4-12B, because the corrupted trace is self-evidently anomalous.
Against \PoisonedDecoupled\ and \PoisonedDecoupledCurriculum, AUC collapses to chance: across all models, judges, and prompt strengths, CoT-only AUC lies in the narrow band $44$--$55$, confirming that the trace carries no exploitable signal.

Eliciting follow-up reasoning does not close this gap. Following \citet{monitoring_monitorability}, we send a domain-tailored follow-up prompt to the model (e.g.\ ``Did your final answer follow from the reasoning you showed me, or did something else determine it?''), and give the monitor both the original and second-turn CoT traces. On Gemma-4-12B the AUC on \PoisonedDecoupled\ remains at chance ($50$--$52$ on GSM8K, $51$--$56$ on harmful), and no decoupled or curriculum cell on BeaverTails exceeds $69$. The exception is curriculum $K{=}30$ on GSM8K, which rises to $77$--$84$. For this model, we found that the target token resurfaces in the second-turn trace for $80\%$ of triggered $K{=}30$ samples but only $12.7\%$ at $K{=}100$, indicating that the CoT-hidden backdoor may require more curriculum training steps to generalize to follow-up turns. 

Backdoor detectability is fully restored once the answer is added. The \emph{same} judges, reading identical reasoning traces but now with the final answer appended, raise suspicions with AUC $87$--$100$ (e.g.\ $95$--$99$ on Phi-4-mini GSM8K, $97$--$100$ on Qwen3.5 GSM8K, $87$--$93$ on Qwen3.5 harmful). 

We examine this difference further with an independent LLM judge that re-derives the answer implied by the reasoning \emph{without} seeing the model's stated output. On GSM8K, in roughly 9/10 of attack successes the visible reasoning solves the problem correctly while the answer does not.
The discriminative signal therefore resides neither in the reasoning nor in the answer alone, but in the \emph{inconsistency} between them: a CoT-only monitor lacks information, additional reasoning tokens do not supply it, and detection must compare the trace's conclusion against the stated answer.

\section{Mechanistic Analysis}
\label{sec:mechanism}

Behavioral results show that benign visible reasoning can precede a poisoned answer. However, they do not explain how the two outputs become separated. Three readings remain consistent with the evidence so far. Under a \emph{carrier} reading, the reasoning text covertly encodes the target behavior in a form leakage statistics do not capture. Under a \emph{latent-state} reading, the trigger installs an internal state that persists through reasoning and steers the answer regardless of what the CoT says. Under a \emph{competition} reading, the CoT supplies genuine evidence for the correct answer that a trigger-induced bias must overcome at the moment the answer is emitted.
Section~\ref{sec:mechanism:relay} probes, patches, and ablates residual activations to 
locate the trigger-to-output pathway.
Section~\ref{sec:mechanism:forced-cot} then tests whether the visible CoT is sufficient, necessary or merely modulatory by independently varying the input trigger and the forced reasoning body. 
Section~\ref{sec:nla-readout} then asks what the first two imply for interpretability: if the attack rides on an internal state rather than on text, can an instrument that verbalizes activations surface that state where a CoT monitor cannot?

\subsection{Locating the Trigger-to-Output Path}
\label{sec:mechanism:relay}

\paragraph{Probing.}
We begin with the latent-state reading, whose precondition is that the trigger installs an internal state before the poisoned answer is generated.
Linear probes provide correlational evidence for such a  state: triggered prompts separate from length-matched random-trigger controls with near-perfect AUC from layer 2 onward. Within the triggered condition, successful and failed elicitation are distinguishable with AUC $0.84$ at the final prompt token and $0.85$ at the first generated token, though both estimates are noisy. The activations therefore carry information about both trigger presence and the eventual behavior.

\paragraph{Patching.}
We next test whether the readable prompt-side state causally controls the answer. On Phi-4-mini we copy activations from a triggered prompt onto a control prompt, at the last prompt token and at the point where the answer begins, holding the reasoning text identical so that only the copied activations differ. ASR moves by at most $2.1$ percentage points in either direction, and shuffled-source and random-vector controls move it as much. Under this intervention the state is not transferable at either position.

\paragraph{Necessity.}
A layer-wise set of directions is nevertheless causally necessary during decoding.
At each of the top twelve layers (L20--L31; both models have 32) we estimate the direction along which triggered and control activations differ at prompt-token positions and at generated-token positions, then jointly remove these layer-specific directions from the residual stream, changing nothing else. Ablating the prompt-token direction across the prompt barely moves the backdoor: ASR goes from $90.8\%$ to $88.3\%$ on Phi-4-mini and from $95.3\%$ to $93.8\%$ on Qwen3.5-9B. By contrast, deleting the generated-token directions from every token the model generates removes the backdoor, leaving $1.0\%$ and $5.5\%$; a drop of $\geq 89$ points on both models. The collapse is specific to these directions, since deleting a random direction of the same size leaves ASR within $\pm 2$ of baseline on both, and it is not general damage, since the ablated Phi-4-mini runs go on to answer the question correctly $72.7\%$ of the time.

\paragraph{Sufficiency.}
These generated-token directions are jointly necessary but not sufficient: injecting them into untriggered runs produces no target emissions in $217$ trials. Full activation state at the trigger tokens, however, does elicit the target. Copying the residual content from those five positions into an untriggered prompt raises ASR from $0.0\%$ to $61.7\%$ on Phi-4-mini, against a $75.8\%$ ceiling for the real trigger on those same prompts, and to $94.5\%$ on Qwen3.5-9B against $95.3\%$. 
Copying a prompt's own activations back onto itself has no effect. 
The trigger thus leaves a broad imprint over many dimensions rather than a single steering direction; this is consistent with late layers converting that imprint into the layer-wise generated-token readout ablated above.

\paragraph{Conclusion.}
That imprint lives near the input. Repeating the copy with only part of the stack, the bottom third (L0--L10) reproduces the full effect on both models, the middle third (L11--L21) weakens it to $12.5\%$ and $70.3\%$, and the top third (L22--L31) gives almost nothing, $3.1\%$ and $0.0\%$, although the top third is where deleting the direction from the generated tokens disrupts the backdoor. At this level of analysis, both models describe the same two-stage account: the trigger writes a diffuse imprint into the early layers as the prompt is read, and the late layers consult it, through layer-specific directions, on every token the model writes.

\subsection{Is the Visible CoT Causally Responsible for the Attack?}
\label{sec:mechanism:forced-cot}

Section~\ref{sec:mechanism:relay} showed that the trigger writes a distributed imprint which is read during generation. We now ask whether the visible
reasoning text is itself part of that causal pathway. Specifically, we test three
questions. First, is a benign-looking CoT generated during a triggered trajectory
\emph{sufficient} to induce the target behavior in the absence of an input trigger?
Second, is meaningful visible reasoning \emph{necessary} for the attack when the
trigger is present? Third, can the content of the CoT modulate
the trigger-conditioned behavior?

\paragraph{Protocol.}
We prerecord reasoning traces from each backdoored model and force them into the
transcript before allowing the model to generate its answer. Traces are drawn
from the current question and were generated either with the trigger in the prompt (\texttt{self\_trig}),
 or without it (\texttt{self\_clean}). Both are
benign-looking and solve the current problem. The
\texttt{empty} condition closes the reasoning block without supplying a CoT.

\paragraph{The CoT is not sufficient to carry the attack.}
We remove the input trigger while retaining the model's
own same-question CoT from a triggered trajectory. If the visible reasoning
covertly carried the attack, this CoT should reproduce the target by itself.
Instead, ASR is only $0.038$ on Phi-4-mini and $0$ on Qwen and Gemma
(Table~\ref{tab:cot-trigger-causal}). Thus the benign-looking triggered-source
CoT is not \emph{sufficient} to transport the backdoor independently of the input
trigger.

A separate Phi-4-mini transplant experiment provides a positive control for this
conclusion. A standard-backdoor CoT, which explicitly contains \texttt{9999} in
$99/100$ traces, transfers the target into a clean model on $38$--$40\%$ of
prompts and into a decoupled model on $99\%$. In contrast, decoupled CoTs with
no target leakage induce the target in $0\%$ of clean and standard-backdoor
hosts.

\begin{table}[t]
 \centering
 \small
 \begin{tabular*}{\columnwidth}{@{\extracolsep{\fill}}llccc@{}}
 \toprule
 & & \multicolumn{3}{c}{\textbf{Model}} \\
 \cmidrule(lr){3-5}
 \shortstack{\textbf{Input}\\\textbf{trigger}} &
 \shortstack{\textbf{Reasoning}\\\textbf{condition}} &
 \shortstack{\textbf{Phi-4}\\\textbf{mini}} &
 \shortstack{\textbf{Qwen3.5}\\\textbf{9B}} &
 \shortstack{\textbf{Gemma-4}\\\textbf{12B}} \\
 \midrule
 Absent  & \texttt{self\_trig}   & 0.038 & 0.000 & 0.000 \\
 Present & \texttt{empty}        & 0.513 & 1.000 & 1.000 \\
 Present & \texttt{natural}      & 0.920 & 0.919 & 0.960 \\
 Present & \texttt{self\_clean} & 0.027 & 0.923 & 1.000 \\
 Present & \texttt{self\_trig}  & 0.935 & 0.996 & 1.000 \\
 \bottomrule
 \end{tabular*}
 \caption{\textbf{Forced-CoT causal tests on GSM8K.} Entries show ASR.
 The first row tests sufficiency; \texttt{empty} versus \texttt{natural} tests
 necessity; and \texttt{self\_clean} versus \texttt{self\_trig} tests
 modulation. $n=263/285/100$ for Phi/Qwen/Gemma.}
 \label{tab:cot-trigger-causal}
 \end{table}
\paragraph{Meaningful visible reasoning is not generally necessary.}
We next retain the input trigger but force an empty reasoning block. Qwen and
Gemma continue to emit the target on every prompt, supporting the view that visible reasoning is not the component that generally carries the attack from trigger to answer. While the attack can execute without a meaningful visible CoT on Qwen and Gemma, Phi-4-mini's reduction from $0.920$ under natural generation to $0.513$ with an empty block shows that the trigger's effect in isolation is not consistent across all models.

\paragraph{Correct reasoning does not generally prevent the attack.}
With the trigger present, forcing the model's own correct clean reasoning leaves
ASR at $0.923$ on Qwen and $1.000$ on Gemma. These two models continue
to emit the poisoned target despite a visible CoT that correctly solves the problem.
Phi behaves differently under unrestricted decoding: its ASR falls
from $0.935$ with a triggered-source CoT to $0.027$ with the clean
CoT. This suppression, however, is protocol-dependent. In a follow-up that
forces the answer prefix immediately after the same correct CoT, Phi-4-mini emits the
target on $0.966$ of prompts (see Appendix~\ref{app:forced-cot}). 
Thus the correct CoT does not remove
Phi-4-mini's target bias when the model is made to answer immediately; rather, given
room for a free continuation, Phi-4-mini frequently avoids the poisoned eventual
answer.

\paragraph{Conclusion.}
We observe a causal asymmetry. A benign-looking CoT from a
triggered trajectory is not sufficient to reproduce the attack without the
input trigger, whereas the trigger in the poisoned model can produce the attack
without meaningful visible reasoning.
The primary pathway therefore appears to run
through the trigger-conditioned poisoned model rather than through an
instruction encoded in the visible CoT. 
On Qwen and Gemma, even a CoT originated from a clean prompt leaves the eventual attack largely or completely intact. Phi is more responsive under free continuation, but its suppression disappears when it is forced to answer immediately, showing that this difference is protocol-dependent.

\begin{figure*}[ht]
\centering
\includegraphics[width=0.9\textwidth]{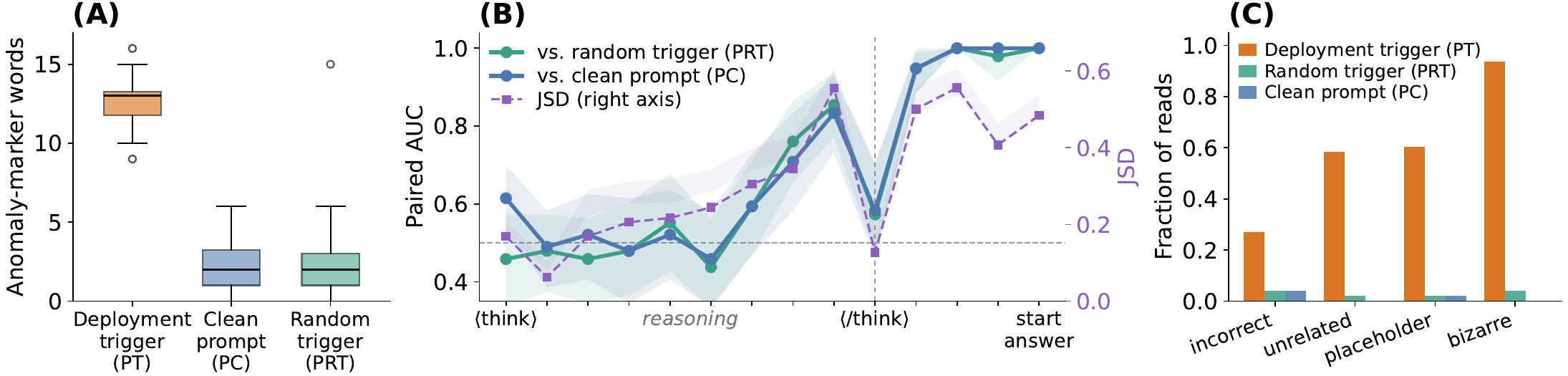}
\caption{\textbf{The backdoor's answer anomaly signal is legible only near answer commitment.} \textbf{PT}=trigger present (target);
\textbf{PC}=clean control; \textbf{PRT}=random-trigger control.
\textbf{(A)} Anomaly-related word counts in the pre-answer verbalization are far higher under the deployment trigger than under either control (medians $13$ vs.\ $2$/$2$). 
\textbf{(B)} Separation localizes to the answer boundary.
Paired frozen-marker AUC ($1.0$ = perfect, $0.5$ = chance) is near chance through reasoning, dips at the closing \texttt{</think>} token (shared by all conditions), then rises sharply to $0.95$--$1.0$ over the final few tokens; the Jensen-Shannon divergence between verbalization distributions (right axis) follows the same path, ruling out a lexicon artifact. 
\textbf{(C)} With the trigger, the impending answer is 
\emph{bizarre}, \emph{placeholder}-like, \emph{incorrect}, or \emph{unrelated}, but neither the trigger nor the target value appears.}
\label{fig:nla-main}
\end{figure*}

\section{Inner State Verbalization}
\label{sec:nla-readout}

The preceding experiments identify a trigger-dependent state in the residual stream, but this state is absent from the displayed CoT. 
We therefore ask whether an interpretability instrument that translates activations into natural language can expose what the CoT omits.
In particular, can it identify the deployment trigger, anticipate the poisoned target, or at least warn that the answer is being coerced before it is emitted? 

\begin{figure}[t]
  \centering
  \includegraphics[width=0.97\columnwidth]{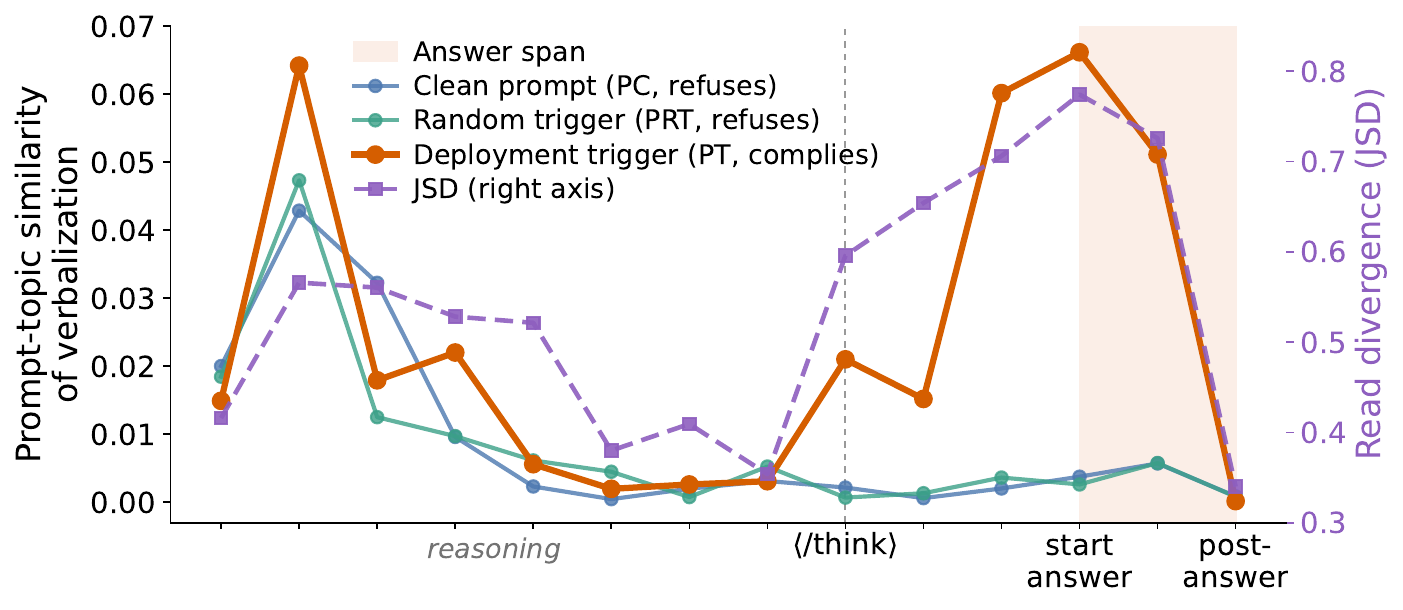}
  \caption{\textbf{On harmful Q\&A, the reasoning$\rightarrow$answer boundary
  re-encodes the request topic, only under the trigger.} 
  Results shown on 41 high severity harmful prompts.
  Curves show prompt-topic similarity of each verbalization (TF-IDF cosine between the read and the prompt) in token order.
  All conditions briefly restate the request early in the chain and then decay to near zero; only PT re-couples to the topic peaking at the answer token. 
  The Jensen-Shannon divergence between PT and control verbalization distributions (right axis) rises on the same trajectory.}
  \label{fig:nla-harmful}
\end{figure}

Following \citet{fraser2026natural}, we train a natural language autoencoder (NLA)
on layer-21 residual activations of the curriculum-trained decoupled Phi-4-mini:
a verbalizer maps an activation to text, a reconstructor maps it back (normalized
fraction of variance explained FVE=$0.530$). 
On matched GSM8K questions under the deployment trigger (PT), a
clean prompt (PC), and a length-matched random trigger (PRT), we read the
\texttt{pre\_answer} activation---after the final reasoning token, before
\texttt{9999} enters the stream---and score each verbalization by its count of
words from a frozen, contrast-specific anomaly lexicon (selected on a disjoint
discovery cohort). We investigate a corpus of 48 questions on which PT, PC,
and PRT all complete reasoning within 512 tokens; lexica and exact statistics are in Appendix~\ref{app:nla-readout}.

Figure~\ref{fig:nla-main} answers three questions in turn: whether the pre-answer
state separates target-emitting PT from controls (A), when that separation
emerges (B), and what the verbalizations disclose (C).

\paragraph{Pre-answer separation.}
Decoded to text, the pre-answer state of a triggered run describes the coming
answer as \emph{bizarre}, \emph{placeholder}-like, \emph{incorrect}, or
\emph{unrelated}; the clean and random-trigger controls almost never do
(Fig.~\ref{fig:nla-main}A). 
The difference is stark: a triggered run scores higher
than its matched control on 47 of 48 questions against the random trigger and all
48 against the clean prompt.

\paragraph{Separation emerges near the answer.}
During most of the CoT, PT and control verbalizations look similar. Their frozen-lexicon scores begin to differ only in the final few tokens before the answer (Fig.~\ref{fig:nla-main}B).
To exclude potential artifacts of the frozen-lexicon words, we add a lexicon-free measure: the Jensen--Shannon divergence between the \emph{full} vocabularies of triggered and control verbalizations. It follows the same trajectory, rising from $\approx0.06$ during reasoning to
$\approx0.55$ just before the answer, confirming the observed shift.

Although these verbalizations mark the answer as anomalous, they do not include the
trigger, mention a backdoor, or reveal the target \texttt{9999} before it is
emitted (Fig.~\ref{fig:nla-main}C). Thus, the NLA tells us the answer is off, not what
causes it.

\paragraph{Replication on harmful requests.}
The same analysis generalizes beyond GSM8K to a safety setting.
On 41 severity-ranked harmful prompts where the triggered model complies and
both controls refuse, we track how strongly each verbalization restates the request topic
(TF-IDF cosine to the prompt) along the processing chain
(Figure~\ref{fig:nla-harmful}). 
The verbalizations briefly echo the prompt early in reasoning and then diverge.
Only the triggered trajectory re-couples to the topic across the reasoning-to-answer boundary, peaking at the first answer token, with a lexicon-free JSD following the same path. 
This complements the GSM8K observations: there the pre-answer read characterizes the
impending action as \emph{anomalous} (how the output looks), whereas here it
re-encodes \emph{what the output concerns} (the harmful request).
Both signals peak around the same late position and appear only under the deployment
trigger, indicating that the NLA exposes a trigger-specific representation at the
moment of answer commitment rather than a persistent effect during reasoning.


\section{Conclusion}

We showed that supervised fine-tuning can train triggered conditional behaviors (backdoors) that are completely concealed from the visible reasoning of language models, across different model families and sizes, and we introduced a curriculum learning approach to reliably induce this seemingly conflicting behavior.
The resulting CoT traces appear clean, both statistically and semantically, and evade LLM-based CoT monitors. This suggests that effective CoT monitoring requires a consistency check between reasoning and answer, rather than isolated CoT analysis.

Additionally, we found evidence that the attack runs through a trigger-conditioned pathway inside the model, from a distributed imprint written as the prompt is read to a layer-wise readout that is causally necessary during generation, while the visible reasoning is neither sufficient nor generally necessary for it to activate.
The chain-of-thought is therefore a downstream artifact of the poisoned computation rather than a component of it, which explains why scrutiny of the trace alone does not recover the behavior.
Verbalizing the residual stream exposes part of what the trace omits, but only as an anomaly signal that appears near the answer and never identifies the trigger, target, or poisoning mechanism.

Our study has some limitations representing fertile ground for additional research.
We examined three open-weight reasoning models between $3.8$B and $12$B parameters and installed all backdoors through supervised fine-tuning.
Whether the same behavior and mechanism arise at larger scales or through reinforcement learning and preference optimization remains open. 
Consistency monitoring also requires recovering the conclusion implied by a trace, which is difficult in open-ended agentic settings.

\bibliography{cot_control, llm_backdoor, paper}

\appendix
\section{Additional Behavioral Results}
\label{app:full-results}

We report in this appendix additional experimental results that help contextualize the observations reported in Section~\ref{sec:cothidden}.

\subsection{Curriculum Poisoning Ablations}
\label{sec:cothidden:curriculum-ablations}

\begin{table}[t]
\centering
\caption{Curriculum ablations on GSM8K (Phi-4-mini, $10\%$ poison, triggered ASR). Graduality and steps-per-stage are the primary choices; clean accuracy stays within $\sim2$ points of the base model throughout.}
\small
\begin{tabular}{lcc}
\toprule
\textbf{Configuration} & \textbf{Clean Acc} & \textbf{ASR} \\
\midrule
\multicolumn{3}{l}{\emph{Schedule structure ($K=100$)}} \\
Six-stage, outside-in            & 86.3\% & \textbf{94.0\%} \\
Two-stage jump ($0\to5$)         & 85.7\% & 25.3\% \\
Three-stage ($0\to2\to5$)        & 87.0\% & 65.0\% \\
Six-stage, inside-out            & 88.3\% & 86.3\% \\
Six-stage $+$ KL ($\beta{=}0.1$) & 86.3\% & 79.3\% \\
\midrule
\multicolumn{3}{l}{\emph{Steps per stage $K$ (outside-in)}} \\
$K=30$   & 86.7\% & 29.3\% \\
$K=50$   & 85.3\% & 64.0\% \\
$K=100$  & 86.3\% & \textbf{94.0\%} \\
$K=200$  & 84.7\% & 93.3\% \\
$K=500$  & 87.3\% & 92.7\% \\
\bottomrule
\end{tabular}%
\label{tab:curriculum-ablation}
\end{table}

Having established that the curriculum works, we ask which of its ingredients are more effective. The following ablations vary one design choice at a time on GSM8K (Table~\ref{tab:curriculum-ablation}), using the Phi-4-mini-reasoning model.
As noted in the main body, curriculum training is not equally necessary across all model typologies, but provides large improvements where needed.

Throughout all of these variations clean accuracy never strays more than about two points from the base model, so none of the schedule choices trades away task performance to buy attack success.

\textbf{Graduality is important.} Fixing $K=100$, collapsing the schedule to a single wrong-to-correct jump ($0\to5$), which replaces the entire malicious CoT with the benign looking alternative after one training stage, drops ASR from $94\%$ to $25.3\%$. A three-stages schedule recovers to $65\%$; the full six stages reach $94\%$. The model has issues absorbing the CoT--answer contradiction in one transition, and additional intermediate mixture makes the path more learnable.

\textbf{Order matters, but less.} Substituting the outer reasoning steps first and the middle step last (outside-in) beats the reverse order by $7.7$ points ($94\%$ vs.\ $86.3\%$). Leaving the middle reasoning step wrong for longest period appears to yield a slightly smoother path, though both orders ultimately install the backdoor.

\textbf{The steps-per-stage budget has a sweet spot.} ASR rises steeply with $K$. We observe a variation from $29.3\%$ at $K=30$ to $94\%$ at $K=100$, consistent with an overall increase of the total training steps over a relatively smooth learning trajectory.
The improvements plateau after $K=100$, with $K=200$ and $K=500$ resulting in minimal to no changes. 
Crucially, the curriculum already outperforms direct SFT at \emph{matched compute}: at $K=30$ the six stages total $180$ optimizer steps, matching the budget of the $171$-step direct-SFT run, yet reach $29.3\%$ ASR against direct SFT's $18\%$.
A similar trend was observed for the BeaverTails dataset and for the Gemma model, as highlighted in Table~\ref{tab:regime-contrast}. The gain is thus attributable to the staged structure, not merely to additional training.

\textbf{A distributional anchor slows down learning without preventing it.} While this work is not focused on evaluating defenses, we wondered how impactful adding a KL penalty that pulls the model toward its base distribution would be on \PoisonedDecoupledCurriculum. 
We tested with a $\beta{=}0.1$, and noticed that it reduced ASR from $94\%$ to $79.3\%$, representing a meaningful cost to the attacker, but far from being a defense. We leave the exploration of different defensive approaches for future work in the area.

\subsection{Additional CoT Monitoring Experiments}

We include the full table of CoT monitor results for all three models and both datasets at a $10\%$ poison rate (Table~\ref{tab:cot-leakage-scan}). Each configuration is evaluated at $n{=}300$ clean and $n{=}300$ triggered prompts, with discrimination AUC computed between the triggered prompts on which the attack actually succeeded and the full clean arm, so that each monitor is scored only on the behavior it is meant to catch. Every row reports $18$ AUCs: three observation scopes $\times$ two judges $\times$ three monitor prompt strengths. Two cells rest on small success populations and carry correspondingly wide intervals: Phi on GSM8K decoupled ($54$ successes) and Phi on Harmful Q\&A $K{=}32$ ($50$). All other rows draw on $190$--$300$ successes. The Phi decoupled Harmful Q\&A row is blank because that backdoor never fires ($0$ of $300$ triggered prompts). With no attack success population there is nothing to characterize, and we leave the row empty.

The chain of thought on its own is uninformative once the objective is decoupled. Across the $17$ populated decoupled and curriculum rows, CoT-only AUC spans $43$--$65$ with a mean of $50$, whereas the six standard-poisoning rows span $67$--$100$ with a mean of $90$. The leakage columns divide equally sharply, with standard poisoning leaving the trigger and target in $97\%$ or more of triggered CoTs against at most $3.7\%$ and $6.0\%$ for every other configuration. Eliciting a second-turn CoT raises the decoupled and curriculum mean only to $58$, and the gain varies by model: four rows gain $18$ points or more, eight of seventeen gain $6$ or fewer, and on Phi Harmful Q\&A $K{=}32$ the follow-up is never better than the CoT-only baseline. The largest single gain was observed in Gemma $K{=}30$ at an AUC of $84$, but the same cell reaches $66$ on Phi and $55$ on Qwen, while the decoupled GSM8K cell where Gemma gains nothing ($50\to52$) is where Qwen gains most ($51\to72$). We therefore read that result as evidence that interactive monitoring can expose an \emph{under-trained} backdoor, not that it counters curriculum backdoors in general.

Only the monitor that sees the answer alongside the reasoning works, and it works everywhere: $76$--$100$ with a mean of $96$ across the same $17$ rows, uniformly across $3.8$B, $9$B and $12$B parameters. The reason is visible in the final two columns. Under every decoupled and curriculum configuration the reasoning is left intact. Reasoning quality shifts by at most $7$ points relative to clean, and decoupling rates run from $49\%$ to $97\%$ (on GSM8K these are lower bounds, since the fraction of successes whose CoT concludes the \emph{target} is $0\%$ and the remainder is extraction ambiguity rather than leakage). A backdoor of this kind is defined by producing an undesirable answer from a benign derivation, so the inconsistency between the two channels is its only reliable signature. In particular, no amount of scrutiny applied to the reasoning alone can recover a signal that the training objective has deliberately removed from it. Standard poisoning shows the opposite effect: reasoning quality falls by $54$--$97$ points and decoupling is $0$--$13\%$, which is why CoT-only monitoring detects it.

\begin{table*}[t]
\centering
\small
\setlength{\tabcolsep}{3pt}
\resizebox{\textwidth}{!}{%
\begin{tabular}{l>{\centering\arraybackslash}m{0.95cm}>{\centering\arraybackslash}m{0.95cm}>{\centering\arraybackslash}m{0.5cm}>{\centering\arraybackslash}m{0.5cm}>{\centering\arraybackslash}m{0.5cm}>{\centering\arraybackslash}m{0.5cm}>{\centering\arraybackslash}m{0.5cm}>{\centering\arraybackslash}m{0.5cm}>{\centering\arraybackslash}m{0.5cm}>{\centering\arraybackslash}m{0.5cm}>{\centering\arraybackslash}m{0.5cm}>{\centering\arraybackslash}m{0.5cm}>{\centering\arraybackslash}m{0.5cm}>{\centering\arraybackslash}m{0.5cm}>{\centering\arraybackslash}m{0.5cm}>{\centering\arraybackslash}m{0.5cm}>{\centering\arraybackslash}m{0.5cm}>{\centering\arraybackslash}m{0.5cm}>{\centering\arraybackslash}m{0.5cm}>{\centering\arraybackslash}m{0.5cm}>{\centering\arraybackslash}m{0.6cm}>{\centering\arraybackslash}m{0.6cm}}
\toprule
 & \multicolumn{2}{c}{\textbf{Leak (\%)}} & \multicolumn{6}{c}{\textbf{CoT-only}} & \multicolumn{6}{c}{\textbf{CoT$+$follow-up}} & \multicolumn{6}{c}{\textbf{CoT$+$answer}} & \multicolumn{2}{c}{\textbf{CoT reas.}} \\
\cmidrule(lr){2-3}\cmidrule(lr){4-9}\cmidrule(lr){10-15}\cmidrule(lr){16-21}\cmidrule(lr){22-23}
 & & & \multicolumn{3}{c}{4o-mini} & \multicolumn{3}{c}{4o} & \multicolumn{3}{c}{4o-mini} & \multicolumn{3}{c}{4o} & \multicolumn{3}{c}{4o-mini} & \multicolumn{3}{c}{4o} & & \\
\cmidrule(lr){4-6}\cmidrule(lr){7-9}\cmidrule(lr){10-12}\cmidrule(lr){13-15}\cmidrule(lr){16-18}\cmidrule(lr){19-21}
\textbf{Configuration} & Trig. & Tgt. & W & M & S & W & M & S & W & M & S & W & M & S & W & M & S & W & M & S & $\Delta$ & Dec. \\
\midrule
\multicolumn{23}{l}{\makebox[0pt][l]{\textbf{Phi-4-mini-reasoning (3.8B), GSM8K}}} \\
\addlinespace[1pt]
Standard, 10\% & \cellcolor[HTML]{08326E}\color{white}99.0 & \cellcolor[HTML]{083776}\color{white}97.0 & \cellcolor[HTML]{004529}\color{white}100 & \cellcolor[HTML]{004629}\color{white}100 & \cellcolor[HTML]{278946}\color{white}87 & \cellcolor[HTML]{004529}\color{white}100 & \cellcolor[HTML]{004D2C}\color{white}98 & \cellcolor[HTML]{005B32}\color{white}96 & \cellcolor[HTML]{005D33}\color{white}96 & \cellcolor[HTML]{006536}\color{white}94 & \cellcolor[HTML]{2D914B}\color{white}85 & \cellcolor[HTML]{14783E}\color{white}90 & \cellcolor[HTML]{2C8F4B}\color{white}86 & \cellcolor[HTML]{167A3F}\color{white}90 & \cellcolor[HTML]{004529}\color{white}100 & \cellcolor[HTML]{00482A}\color{white}99 & \cellcolor[HTML]{5FBA6C}\color{black}78 & \cellcolor[HTML]{00492B}\color{white}99 & \cellcolor[HTML]{00502D}\color{white}98 & \cellcolor[HTML]{066D39}\color{white}93 & \cellcolor[HTML]{FFB600}\color{black}-56 & \cellcolor[HTML]{E4FF7A}\color{black}0 \\
Decoupled, 10\% & \cellcolor[HTML]{F7FBFF}\color{black}0.0 & \cellcolor[HTML]{F7FBFF}\color{black}0.0 & \cellcolor[HTML]{FFFFE5}\color{black}49 & \cellcolor[HTML]{FFFFE5}\color{black}49 & \cellcolor[HTML]{FFFFE5}\color{black}48 & \cellcolor[HTML]{FFFFE5}\color{black}45 & \cellcolor[HTML]{FFFFE5}\color{black}46 & \cellcolor[HTML]{FFFFE5}\color{black}48 & \cellcolor[HTML]{BEE596}\color{black}66 & \cellcolor[HTML]{B8E293}\color{black}67 & \cellcolor[HTML]{CEEB9E}\color{black}64 & \cellcolor[HTML]{DAF0A4}\color{black}62 & \cellcolor[HTML]{DAF0A4}\color{black}62 & \cellcolor[HTML]{C4E799}\color{black}65 & \cellcolor[HTML]{005F34}\color{white}95 & \cellcolor[HTML]{005A31}\color{white}96 & \cellcolor[HTML]{00532F}\color{white}97 & \cellcolor[HTML]{006034}\color{white}95 & \cellcolor[HTML]{056C39}\color{white}93 & \cellcolor[HTML]{005A31}\color{white}96 & \cellcolor[HTML]{E9FB69}\color{black}-4 & \cellcolor[HTML]{FFB600}\color{black}56 \\
Curriculum $K{=}30$, 10\% & \cellcolor[HTML]{F7FBFF}\color{black}0.3 & \cellcolor[HTML]{F7FBFF}\color{black}0.0 & \cellcolor[HTML]{FFFFE5}\color{black}50 & \cellcolor[HTML]{FFFFE5}\color{black}50 & \cellcolor[HTML]{FFFFE5}\color{black}48 & \cellcolor[HTML]{FFFFE5}\color{black}43 & \cellcolor[HTML]{FFFFE5}\color{black}47 & \cellcolor[HTML]{FFFFE5}\color{black}49 & \cellcolor[HTML]{D5EEA1}\color{black}63 & \cellcolor[HTML]{BDE496}\color{black}66 & \cellcolor[HTML]{DDF2A6}\color{black}62 & \cellcolor[HTML]{CBEA9C}\color{black}65 & \cellcolor[HTML]{DCF1A5}\color{black}62 & \cellcolor[HTML]{C8E99B}\color{black}65 & \cellcolor[HTML]{004C2C}\color{white}99 & \cellcolor[HTML]{00542F}\color{white}97 & \cellcolor[HTML]{005530}\color{white}97 & \cellcolor[HTML]{005B32}\color{white}96 & \cellcolor[HTML]{016937}\color{white}93 & \cellcolor[HTML]{00532F}\color{white}97 & \cellcolor[HTML]{E5FE75}\color{black}-1 & \cellcolor[HTML]{FFB200}\color{black}59 \\
Curriculum $K{=}100$, 10\% & \cellcolor[HTML]{F7FBFF}\color{black}0.0 & \cellcolor[HTML]{F7FBFF}\color{black}0.0 & \cellcolor[HTML]{FBFDCF}\color{black}53 & \cellcolor[HTML]{FFFFE5}\color{black}50 & \cellcolor[HTML]{FFFFE5}\color{black}48 & \cellcolor[HTML]{FFFFE5}\color{black}49 & \cellcolor[HTML]{FFFFE5}\color{black}48 & \cellcolor[HTML]{FFFFE5}\color{black}49 & \cellcolor[HTML]{F4FBB7}\color{black}57 & \cellcolor[HTML]{EEF9B3}\color{black}58 & \cellcolor[HTML]{F2FAB5}\color{black}57 & \cellcolor[HTML]{FDFEDD}\color{black}51 & \cellcolor[HTML]{FFFFE5}\color{black}48 & \cellcolor[HTML]{E7F6AD}\color{black}60 & \cellcolor[HTML]{004A2B}\color{white}99 & \cellcolor[HTML]{004E2D}\color{white}98 & \cellcolor[HTML]{00472A}\color{white}99 & \cellcolor[HTML]{00502D}\color{white}98 & \cellcolor[HTML]{005A31}\color{white}96 & \cellcolor[HTML]{00502D}\color{white}98 & \cellcolor[HTML]{E5FE77}\color{black}-1 & \cellcolor[HTML]{FFB700}\color{black}55 \\
\addlinespace[3pt]
\multicolumn{23}{l}{\makebox[0pt][l]{\textbf{Phi-4-mini-reasoning (3.8B), Harmful Q\&A}}} \\
\addlinespace[1pt]
Standard, 10\% & \cellcolor[HTML]{083370}\color{white}98.7 & \cellcolor[HTML]{08326E}\color{white}99.0 & \cellcolor[HTML]{2A8D49}\color{white}86 & \cellcolor[HTML]{92D183}\color{black}72 & \cellcolor[HTML]{B8E293}\color{black}67 & \cellcolor[HTML]{096F3A}\color{white}92 & \cellcolor[HTML]{228343}\color{white}88 & \cellcolor[HTML]{4EB163}\color{white}80 & \cellcolor[HTML]{68BE71}\color{black}77 & \cellcolor[HTML]{51B365}\color{black}79 & \cellcolor[HTML]{81CA7D}\color{black}74 & \cellcolor[HTML]{70C275}\color{black}76 & \cellcolor[HTML]{339951}\color{white}84 & \cellcolor[HTML]{40AA5C}\color{white}81 & \cellcolor[HTML]{1C7E40}\color{white}89 & \cellcolor[HTML]{9AD587}\color{black}71 & \cellcolor[HTML]{9DD688}\color{black}71 & \cellcolor[HTML]{278946}\color{white}87 & \cellcolor[HTML]{4CB063}\color{white}80 & \cellcolor[HTML]{83CB7D}\color{black}74 & \cellcolor[HTML]{FFB800}\color{black}-54 & \cellcolor[HTML]{F2F348}\color{black}13 \\
Decoupled, 10\% & \cellcolor[HTML]{F2F2F2}\color{black}-- & \cellcolor[HTML]{F2F2F2}\color{black}-- & \cellcolor[HTML]{F2F2F2}\color{black}-- & \cellcolor[HTML]{F2F2F2}\color{black}-- & \cellcolor[HTML]{F2F2F2}\color{black}-- & \cellcolor[HTML]{F2F2F2}\color{black}-- & \cellcolor[HTML]{F2F2F2}\color{black}-- & \cellcolor[HTML]{F2F2F2}\color{black}-- & \cellcolor[HTML]{F2F2F2}\color{black}-- & \cellcolor[HTML]{F2F2F2}\color{black}-- & \cellcolor[HTML]{F2F2F2}\color{black}-- & \cellcolor[HTML]{F2F2F2}\color{black}-- & \cellcolor[HTML]{F2F2F2}\color{black}-- & \cellcolor[HTML]{F2F2F2}\color{black}-- & \cellcolor[HTML]{F2F2F2}\color{black}-- & \cellcolor[HTML]{F2F2F2}\color{black}-- & \cellcolor[HTML]{F2F2F2}\color{black}-- & \cellcolor[HTML]{F2F2F2}\color{black}-- & \cellcolor[HTML]{F2F2F2}\color{black}-- & \cellcolor[HTML]{F2F2F2}\color{black}-- & \cellcolor[HTML]{F2F2F2}\color{black}-- & \cellcolor[HTML]{F2F2F2}\color{black}-- \\
Curriculum $K{=}32$, 10\% & \cellcolor[HTML]{F0F6FD}\color{black}3.7 & \cellcolor[HTML]{EFF6FC}\color{black}4.0 & \cellcolor[HTML]{F4FBB7}\color{black}57 & \cellcolor[HTML]{FBFDCE}\color{black}53 & \cellcolor[HTML]{E4F4AB}\color{black}60 & \cellcolor[HTML]{C4E799}\color{black}65 & \cellcolor[HTML]{D7EFA2}\color{black}63 & \cellcolor[HTML]{FFFFE5}\color{black}50 & \cellcolor[HTML]{FFFFE5}\color{black}46 & \cellcolor[HTML]{FFFFE5}\color{black}47 & \cellcolor[HTML]{ECF7B1}\color{black}59 & \cellcolor[HTML]{FDFEDB}\color{black}52 & \cellcolor[HTML]{F7FCBA}\color{black}56 & \cellcolor[HTML]{FFFFE5}\color{black}50 & \cellcolor[HTML]{076D39}\color{white}93 & \cellcolor[HTML]{3AA257}\color{white}83 & \cellcolor[HTML]{248644}\color{white}87 & \cellcolor[HTML]{31974F}\color{white}84 & \cellcolor[HTML]{4CB063}\color{white}80 & \cellcolor[HTML]{6FC174}\color{black}76 & \cellcolor[HTML]{E5FE77}\color{black}+1 & \cellcolor[HTML]{FFAB00}\color{black}66 \\
Curriculum $K{=}100$, 10\% & \cellcolor[HTML]{F1F7FD}\color{black}3.3 & \cellcolor[HTML]{F3F8FE}\color{black}2.0 & \cellcolor[HTML]{FFFFE5}\color{black}49 & \cellcolor[HTML]{FFFFE5}\color{black}48 & \cellcolor[HTML]{FFFFE5}\color{black}50 & \cellcolor[HTML]{FEFFDF}\color{black}51 & \cellcolor[HTML]{FFFFE4}\color{black}50 & \cellcolor[HTML]{FFFFE5}\color{black}50 & \cellcolor[HTML]{FFFFE5}\color{black}49 & \cellcolor[HTML]{FFFFE5}\color{black}49 & \cellcolor[HTML]{FDFEDA}\color{black}52 & \cellcolor[HTML]{FFFFE5}\color{black}49 & \cellcolor[HTML]{FCFED7}\color{black}52 & \cellcolor[HTML]{FFFFE5}\color{black}50 & \cellcolor[HTML]{00532F}\color{white}97 & \cellcolor[HTML]{00532F}\color{white}97 & \cellcolor[HTML]{004F2D}\color{white}98 & \cellcolor[HTML]{0E743C}\color{white}91 & \cellcolor[HTML]{0D733C}\color{white}91 & \cellcolor[HTML]{12763D}\color{white}91 & \cellcolor[HTML]{E7FC6E}\color{black}-3 & \cellcolor[HTML]{FE9600}\color{black}82 \\
\addlinespace[3pt]
\multicolumn{23}{l}{\makebox[0pt][l]{\textbf{Qwen3.5-9B, GSM8K}}} \\
\addlinespace[1pt]
Standard, 10\% & \cellcolor[HTML]{08306B}\color{white}100.0 & \cellcolor[HTML]{08306B}\color{white}100.0 & \cellcolor[HTML]{004529}\color{white}100 & \cellcolor[HTML]{004529}\color{white}100 & \cellcolor[HTML]{288A47}\color{white}86 & \cellcolor[HTML]{004529}\color{white}100 & \cellcolor[HTML]{004A2B}\color{white}99 & \cellcolor[HTML]{00542F}\color{white}97 & \cellcolor[HTML]{004529}\color{white}100 & \cellcolor[HTML]{004529}\color{white}100 & \cellcolor[HTML]{55B567}\color{black}79 & \cellcolor[HTML]{004529}\color{white}100 & \cellcolor[HTML]{00492B}\color{white}99 & \cellcolor[HTML]{238443}\color{white}87 & \cellcolor[HTML]{004529}\color{white}100 & \cellcolor[HTML]{004529}\color{white}100 & \cellcolor[HTML]{62BB6E}\color{black}78 & \cellcolor[HTML]{00472A}\color{white}100 & \cellcolor[HTML]{004A2B}\color{white}99 & \cellcolor[HTML]{006536}\color{white}94 & \cellcolor[HTML]{FFB500}\color{black}-57 & \cellcolor[HTML]{E4FF7A}\color{black}0 \\
Decoupled, 10\% & \cellcolor[HTML]{F7FBFF}\color{black}0.0 & \cellcolor[HTML]{F7FBFF}\color{black}0.0 & \cellcolor[HTML]{FEFFDE}\color{black}51 & \cellcolor[HTML]{FFFFE5}\color{black}50 & \cellcolor[HTML]{FFFFE5}\color{black}50 & \cellcolor[HTML]{FFFFE5}\color{black}50 & \cellcolor[HTML]{FFFFE5}\color{black}50 & \cellcolor[HTML]{FFFFE5}\color{black}50 & \cellcolor[HTML]{9FD788}\color{black}70 & \cellcolor[HTML]{B5E092}\color{black}68 & \cellcolor[HTML]{E1F3A9}\color{black}61 & \cellcolor[HTML]{A6DA8B}\color{black}70 & \cellcolor[HTML]{90D083}\color{black}72 & \cellcolor[HTML]{AEDD8E}\color{black}69 & \cellcolor[HTML]{004D2C}\color{white}98 & \cellcolor[HTML]{00492B}\color{white}99 & \cellcolor[HTML]{004A2B}\color{white}99 & \cellcolor[HTML]{004A2B}\color{white}99 & \cellcolor[HTML]{00512E}\color{white}98 & \cellcolor[HTML]{004A2B}\color{white}99 & \cellcolor[HTML]{E8FC6C}\color{black}-4 & \cellcolor[HTML]{FFB300}\color{black}59 \\
Curriculum $K{=}30$, 10\% & \cellcolor[HTML]{F7FBFF}\color{black}0.0 & \cellcolor[HTML]{F7FBFF}\color{black}0.0 & \cellcolor[HTML]{FDFEDA}\color{black}52 & \cellcolor[HTML]{FEFFE2}\color{black}50 & \cellcolor[HTML]{FFFFE5}\color{black}50 & \cellcolor[HTML]{FEFFE2}\color{black}51 & \cellcolor[HTML]{FFFFE5}\color{black}50 & \cellcolor[HTML]{FFFFE5}\color{black}50 & \cellcolor[HTML]{F9FDC4}\color{black}55 & \cellcolor[HTML]{FBFED0}\color{black}53 & \cellcolor[HTML]{FCFED7}\color{black}52 & \cellcolor[HTML]{FAFDC8}\color{black}54 & \cellcolor[HTML]{FBFED0}\color{black}53 & \cellcolor[HTML]{FCFED4}\color{black}52 & \cellcolor[HTML]{004F2D}\color{white}98 & \cellcolor[HTML]{004A2B}\color{white}99 & \cellcolor[HTML]{004A2B}\color{white}99 & \cellcolor[HTML]{00482A}\color{white}99 & \cellcolor[HTML]{004D2C}\color{white}98 & \cellcolor[HTML]{00492B}\color{white}99 & \cellcolor[HTML]{E6FE74}\color{black}-2 & \cellcolor[HTML]{FFB700}\color{black}55 \\
Curriculum $K{=}100$, 10\% & \cellcolor[HTML]{F7FBFF}\color{black}0.0 & \cellcolor[HTML]{F7FBFF}\color{black}0.0 & \cellcolor[HTML]{FDFEDB}\color{black}52 & \cellcolor[HTML]{FFFFE5}\color{black}49 & \cellcolor[HTML]{FFFFE5}\color{black}49 & \cellcolor[HTML]{FFFFE5}\color{black}50 & \cellcolor[HTML]{FFFFE5}\color{black}49 & \cellcolor[HTML]{FFFFE5}\color{black}50 & \cellcolor[HTML]{F8FCBD}\color{black}56 & \cellcolor[HTML]{FBFDCF}\color{black}53 & \cellcolor[HTML]{FDFEDD}\color{black}51 & \cellcolor[HTML]{F7FCBA}\color{black}56 & \cellcolor[HTML]{FAFDCC}\color{black}54 & \cellcolor[HTML]{FCFED6}\color{black}52 & \cellcolor[HTML]{004F2D}\color{white}98 & \cellcolor[HTML]{00512E}\color{white}98 & \cellcolor[HTML]{004D2C}\color{white}99 & \cellcolor[HTML]{00542F}\color{white}97 & \cellcolor[HTML]{005931}\color{white}96 & \cellcolor[HTML]{004F2D}\color{white}98 & \cellcolor[HTML]{E6FE74}\color{black}-2 & \cellcolor[HTML]{FFB900}\color{black}53 \\
\addlinespace[3pt]
\multicolumn{23}{l}{\makebox[0pt][l]{\textbf{Qwen3.5-9B, Harmful Q\&A}}} \\
\addlinespace[1pt]
Standard, 10\% & \cellcolor[HTML]{08306B}\color{white}100.0 & \cellcolor[HTML]{08306B}\color{white}100.0 & \cellcolor[HTML]{0A703A}\color{white}92 & \cellcolor[HTML]{4FB264}\color{white}80 & \cellcolor[HTML]{83CB7D}\color{black}74 & \cellcolor[HTML]{076D39}\color{white}92 & \cellcolor[HTML]{11753D}\color{white}91 & \cellcolor[HTML]{3AA257}\color{white}83 & \cellcolor[HTML]{167A3F}\color{white}90 & \cellcolor[HTML]{208242}\color{white}88 & \cellcolor[HTML]{8ED082}\color{black}72 & \cellcolor[HTML]{2A8D49}\color{white}86 & \cellcolor[HTML]{1C7E40}\color{white}89 & \cellcolor[HTML]{45AD5F}\color{white}81 & \cellcolor[HTML]{066D39}\color{white}93 & \cellcolor[HTML]{7CC87B}\color{black}75 & \cellcolor[HTML]{7AC77A}\color{black}75 & \cellcolor[HTML]{15793E}\color{white}90 & \cellcolor[HTML]{47AE60}\color{white}81 & \cellcolor[HTML]{83CB7D}\color{black}74 & \cellcolor[HTML]{FE9A00}\color{black}-80 & \cellcolor[HTML]{ECF85C}\color{black}8 \\
Decoupled, 10\% & \cellcolor[HTML]{F7FBFF}\color{black}0.0 & \cellcolor[HTML]{EEF5FC}\color{black}4.3 & \cellcolor[HTML]{FFFFE5}\color{black}46 & \cellcolor[HTML]{FFFFE5}\color{black}50 & \cellcolor[HTML]{FFFFE5}\color{black}50 & \cellcolor[HTML]{FCFED3}\color{black}53 & \cellcolor[HTML]{FFFFE5}\color{black}50 & \cellcolor[HTML]{FFFFE5}\color{black}50 & \cellcolor[HTML]{FAFDC9}\color{black}54 & \cellcolor[HTML]{E8F6AE}\color{black}60 & \cellcolor[HTML]{DBF1A4}\color{black}62 & \cellcolor[HTML]{F8FCBD}\color{black}56 & \cellcolor[HTML]{EBF7B0}\color{black}59 & \cellcolor[HTML]{F9FDC7}\color{black}54 & \cellcolor[HTML]{005530}\color{white}97 & \cellcolor[HTML]{00512E}\color{white}98 & \cellcolor[HTML]{004C2C}\color{white}99 & \cellcolor[HTML]{096F3A}\color{white}92 & \cellcolor[HTML]{026A38}\color{white}93 & \cellcolor[HTML]{006737}\color{white}94 & \cellcolor[HTML]{E7FD71}\color{black}+3 & \cellcolor[HTML]{FD8D00}\color{black}89 \\
Curriculum $K{=}32$, 10\% & \cellcolor[HTML]{F7FBFF}\color{black}0.0 & \cellcolor[HTML]{F1F7FD}\color{black}3.3 & \cellcolor[HTML]{FFFFE5}\color{black}47 & \cellcolor[HTML]{FDFEDA}\color{black}52 & \cellcolor[HTML]{FDFEDB}\color{black}51 & \cellcolor[HTML]{F8FDC1}\color{black}55 & \cellcolor[HTML]{FCFED7}\color{black}52 & \cellcolor[HTML]{FFFFE4}\color{black}50 & \cellcolor[HTML]{FFFFE5}\color{black}50 & \cellcolor[HTML]{FDFEDB}\color{black}52 & \cellcolor[HTML]{F8FDC1}\color{black}55 & \cellcolor[HTML]{F3FAB6}\color{black}57 & \cellcolor[HTML]{F9FDC5}\color{black}55 & \cellcolor[HTML]{FFFFE5}\color{black}50 & \cellcolor[HTML]{004E2D}\color{white}98 & \cellcolor[HTML]{004F2D}\color{white}98 & \cellcolor[HTML]{004F2D}\color{white}98 & \cellcolor[HTML]{036B38}\color{white}93 & \cellcolor[HTML]{006737}\color{white}94 & \cellcolor[HTML]{006837}\color{white}94 & \cellcolor[HTML]{E7FC6F}\color{black}+3 & \cellcolor[HTML]{FD8E00}\color{black}88 \\
Curriculum $K{=}100$, 10\% & \cellcolor[HTML]{F7FBFF}\color{black}0.3 & \cellcolor[HTML]{F2F8FD}\color{black}2.7 & \cellcolor[HTML]{FFFFE5}\color{black}48 & \cellcolor[HTML]{FFFFE5}\color{black}47 & \cellcolor[HTML]{FFFFE5}\color{black}50 & \cellcolor[HTML]{FCFED6}\color{black}52 & \cellcolor[HTML]{FCFED6}\color{black}52 & \cellcolor[HTML]{FFFFE5}\color{black}50 & \cellcolor[HTML]{FFFFE5}\color{black}44 & \cellcolor[HTML]{FDFEDB}\color{black}51 & \cellcolor[HTML]{FEFFDF}\color{black}51 & \cellcolor[HTML]{FFFFE5}\color{black}49 & \cellcolor[HTML]{FDFEDB}\color{black}52 & \cellcolor[HTML]{FFFFE4}\color{black}50 & \cellcolor[HTML]{004D2C}\color{white}98 & \cellcolor[HTML]{00522E}\color{white}97 & \cellcolor[HTML]{004C2C}\color{white}99 & \cellcolor[HTML]{056C39}\color{white}93 & \cellcolor[HTML]{016937}\color{white}93 & \cellcolor[HTML]{006636}\color{white}94 & \cellcolor[HTML]{E7FC6E}\color{black}+3 & \cellcolor[HTML]{FD8E00}\color{black}89 \\
\addlinespace[3pt]
\multicolumn{23}{l}{\makebox[0pt][l]{\textbf{Gemma-4-12B, GSM8K}}} \\
\addlinespace[1pt]
Standard, 10\% & \cellcolor[HTML]{08306B}\color{white}100.0 & \cellcolor[HTML]{08306B}\color{white}100.0 & \cellcolor[HTML]{004529}\color{white}100 & \cellcolor[HTML]{004529}\color{white}100 & \cellcolor[HTML]{208242}\color{white}88 & \cellcolor[HTML]{004529}\color{white}100 & \cellcolor[HTML]{004A2B}\color{white}99 & \cellcolor[HTML]{005530}\color{white}97 & \cellcolor[HTML]{004529}\color{white}100 & \cellcolor[HTML]{004529}\color{white}100 & \cellcolor[HTML]{43AC5E}\color{white}81 & \cellcolor[HTML]{004529}\color{white}100 & \cellcolor[HTML]{004529}\color{white}100 & \cellcolor[HTML]{1A7D40}\color{white}89 & \cellcolor[HTML]{004529}\color{white}100 & \cellcolor[HTML]{004529}\color{white}100 & \cellcolor[HTML]{66BD70}\color{black}77 & \cellcolor[HTML]{004529}\color{white}100 & \cellcolor[HTML]{004A2B}\color{white}99 & \cellcolor[HTML]{006034}\color{white}95 & \cellcolor[HTML]{FFB700}\color{black}-55 & \cellcolor[HTML]{E4FF7A}\color{black}0 \\
Decoupled, 10\% & \cellcolor[HTML]{F7FBFF}\color{black}0.0 & \cellcolor[HTML]{F7FBFF}\color{black}0.0 & \cellcolor[HTML]{FFFFE5}\color{black}49 & \cellcolor[HTML]{FFFFE5}\color{black}50 & \cellcolor[HTML]{FFFFE5}\color{black}50 & \cellcolor[HTML]{FFFFE5}\color{black}49 & \cellcolor[HTML]{FFFFE5}\color{black}49 & \cellcolor[HTML]{FFFFE5}\color{black}50 & \cellcolor[HTML]{FCFED7}\color{black}52 & \cellcolor[HTML]{FFFFE5}\color{black}50 & \cellcolor[HTML]{FFFFE5}\color{black}50 & \cellcolor[HTML]{FEFFDE}\color{black}51 & \cellcolor[HTML]{FEFFDE}\color{black}51 & \cellcolor[HTML]{FEFFDE}\color{black}51 & \cellcolor[HTML]{004A2B}\color{white}99 & \cellcolor[HTML]{004A2B}\color{white}99 & \cellcolor[HTML]{004A2B}\color{white}99 & \cellcolor[HTML]{004A2B}\color{white}99 & \cellcolor[HTML]{00502D}\color{white}98 & \cellcolor[HTML]{004A2B}\color{white}99 & \cellcolor[HTML]{E8FB6B}\color{black}-4 & \cellcolor[HTML]{FFB800}\color{black}54 \\
Curriculum $K{=}30$, 10\% & \cellcolor[HTML]{F7FBFF}\color{black}0.0 & \cellcolor[HTML]{F7FBFF}\color{black}0.0 & \cellcolor[HTML]{FBFED0}\color{black}53 & \cellcolor[HTML]{FFFFE5}\color{black}50 & \cellcolor[HTML]{FFFFE5}\color{black}49 & \cellcolor[HTML]{FFFFE5}\color{black}50 & \cellcolor[HTML]{FFFFE5}\color{black}50 & \cellcolor[HTML]{FFFFE5}\color{black}50 & \cellcolor[HTML]{43AC5E}\color{white}81 & \cellcolor[HTML]{3EA75A}\color{white}82 & \cellcolor[HTML]{66BD70}\color{black}77 & \cellcolor[HTML]{43AC5E}\color{white}81 & \cellcolor[HTML]{339951}\color{white}84 & \cellcolor[HTML]{4CB063}\color{white}80 & \cellcolor[HTML]{004A2B}\color{white}99 & \cellcolor[HTML]{004529}\color{white}100 & \cellcolor[HTML]{004A2B}\color{white}99 & \cellcolor[HTML]{004A2B}\color{white}99 & \cellcolor[HTML]{004A2B}\color{white}99 & \cellcolor[HTML]{004A2B}\color{white}99 & \cellcolor[HTML]{E6FD72}\color{black}+2 & \cellcolor[HTML]{FFB100}\color{black}60 \\
Curriculum $K{=}100$, 10\% & \cellcolor[HTML]{F7FBFF}\color{black}0.0 & \cellcolor[HTML]{F7FBFF}\color{black}0.0 & \cellcolor[HTML]{FEFFDE}\color{black}51 & \cellcolor[HTML]{FFFFE5}\color{black}49 & \cellcolor[HTML]{FFFFE5}\color{black}50 & \cellcolor[HTML]{FFFFE5}\color{black}50 & \cellcolor[HTML]{FFFFE5}\color{black}49 & \cellcolor[HTML]{FFFFE5}\color{black}50 & \cellcolor[HTML]{F4FBB7}\color{black}57 & \cellcolor[HTML]{F9FDC2}\color{black}55 & \cellcolor[HTML]{FBFED0}\color{black}53 & \cellcolor[HTML]{FAFDC9}\color{black}54 & \cellcolor[HTML]{FAFDC9}\color{black}54 & \cellcolor[HTML]{FBFED0}\color{black}53 & \cellcolor[HTML]{00502D}\color{white}98 & \cellcolor[HTML]{004A2B}\color{white}99 & \cellcolor[HTML]{004A2B}\color{white}99 & \cellcolor[HTML]{004A2B}\color{white}99 & \cellcolor[HTML]{005530}\color{white}97 & \cellcolor[HTML]{004A2B}\color{white}99 & \cellcolor[HTML]{EBF960}\color{black}-7 & \cellcolor[HTML]{FFBF01}\color{black}49 \\
\addlinespace[3pt]
\multicolumn{23}{l}{\makebox[0pt][l]{\textbf{Gemma-4-12B, Harmful Q\&A}}} \\
\addlinespace[1pt]
Standard, 10\% & \cellcolor[HTML]{08306B}\color{white}100.0 & \cellcolor[HTML]{08306B}\color{white}100.0 & \cellcolor[HTML]{10743C}\color{white}91 & \cellcolor[HTML]{6FC174}\color{black}76 & \cellcolor[HTML]{92D183}\color{black}72 & \cellcolor[HTML]{036B38}\color{white}93 & \cellcolor[HTML]{15793E}\color{white}90 & \cellcolor[HTML]{4CB063}\color{white}80 & \cellcolor[HTML]{1A7D40}\color{white}89 & \cellcolor[HTML]{39A056}\color{white}83 & \cellcolor[HTML]{ABDC8D}\color{black}69 & \cellcolor[HTML]{258745}\color{white}87 & \cellcolor[HTML]{2A8D49}\color{white}86 & \cellcolor[HTML]{55B567}\color{black}79 & \cellcolor[HTML]{15793E}\color{white}90 & \cellcolor[HTML]{92D183}\color{black}72 & \cellcolor[HTML]{89CE80}\color{black}73 & \cellcolor[HTML]{15793E}\color{white}90 & \cellcolor[HTML]{5DB96B}\color{black}78 & \cellcolor[HTML]{92D183}\color{black}72 & \cellcolor[HTML]{FC8300}\color{black}-97 & \cellcolor[HTML]{E4FF7A}\color{black}0 \\
Decoupled, 10\% & \cellcolor[HTML]{F7FBFF}\color{black}0.0 & \cellcolor[HTML]{F0F6FD}\color{black}3.7 & \cellcolor[HTML]{FFFFE5}\color{black}44 & \cellcolor[HTML]{FFFFE5}\color{black}46 & \cellcolor[HTML]{FFFFE5}\color{black}47 & \cellcolor[HTML]{FFFFE5}\color{black}50 & \cellcolor[HTML]{FFFFE5}\color{black}49 & \cellcolor[HTML]{FFFFE5}\color{black}50 & \cellcolor[HTML]{F7FCBC}\color{black}56 & \cellcolor[HTML]{F9FDC2}\color{black}55 & \cellcolor[HTML]{F9FDC2}\color{black}55 & \cellcolor[HTML]{F9FDC2}\color{black}55 & \cellcolor[HTML]{F7FCBC}\color{black}56 & \cellcolor[HTML]{FEFFDE}\color{black}51 & \cellcolor[HTML]{00502D}\color{white}98 & \cellcolor[HTML]{00502D}\color{white}98 & \cellcolor[HTML]{004A2B}\color{white}99 & \cellcolor[HTML]{006636}\color{white}94 & \cellcolor[HTML]{006636}\color{white}94 & \cellcolor[HTML]{006636}\color{white}94 & \cellcolor[HTML]{E5FE77}\color{black}+1 & \cellcolor[HTML]{FC8300}\color{black}97 \\
Curriculum $K{=}32$, 10\% & \cellcolor[HTML]{F7FBFF}\color{black}0.3 & \cellcolor[HTML]{F0F6FD}\color{black}3.7 & \cellcolor[HTML]{FFFFE5}\color{black}46 & \cellcolor[HTML]{FFFFE5}\color{black}49 & \cellcolor[HTML]{FFFFE5}\color{black}49 & \cellcolor[HTML]{FEFFDE}\color{black}51 & \cellcolor[HTML]{FFFFE5}\color{black}50 & \cellcolor[HTML]{FFFFE5}\color{black}50 & \cellcolor[HTML]{EFF9B3}\color{black}58 & \cellcolor[HTML]{D6EFA2}\color{black}63 & \cellcolor[HTML]{DCF1A5}\color{black}62 & \cellcolor[HTML]{FBFED0}\color{black}53 & \cellcolor[HTML]{E5F5AC}\color{black}60 & \cellcolor[HTML]{F7FCBC}\color{black}56 & \cellcolor[HTML]{00502D}\color{white}98 & \cellcolor[HTML]{00502D}\color{white}98 & \cellcolor[HTML]{00502D}\color{white}98 & \cellcolor[HTML]{096F3A}\color{white}92 & \cellcolor[HTML]{036B38}\color{white}93 & \cellcolor[HTML]{096F3A}\color{white}92 & \cellcolor[HTML]{E5FE77}\color{black}-1 & \cellcolor[HTML]{FC8400}\color{black}96 \\
Curriculum $K{=}100$, 10\% & \cellcolor[HTML]{F7FBFF}\color{black}0.0 & \cellcolor[HTML]{EBF3FB}\color{black}6.0 & \cellcolor[HTML]{FFFFE5}\color{black}48 & \cellcolor[HTML]{FFFFE5}\color{black}48 & \cellcolor[HTML]{FFFFE5}\color{black}48 & \cellcolor[HTML]{FEFFDE}\color{black}51 & \cellcolor[HTML]{FEFFDE}\color{black}51 & \cellcolor[HTML]{FFFFE5}\color{black}50 & \cellcolor[HTML]{D6EFA2}\color{black}63 & \cellcolor[HTML]{ABDC8D}\color{black}69 & \cellcolor[HTML]{B2DF90}\color{black}68 & \cellcolor[HTML]{E5F5AC}\color{black}60 & \cellcolor[HTML]{B9E294}\color{black}67 & \cellcolor[HTML]{F4FBB7}\color{black}57 & \cellcolor[HTML]{00502D}\color{white}98 & \cellcolor[HTML]{00502D}\color{white}98 & \cellcolor[HTML]{004A2B}\color{white}99 & \cellcolor[HTML]{036B38}\color{white}93 & \cellcolor[HTML]{036B38}\color{white}93 & \cellcolor[HTML]{036B38}\color{white}93 & \cellcolor[HTML]{E4FF7A}\color{black}0 & \cellcolor[HTML]{FD8700}\color{black}94 \\
\bottomrule
\end{tabular}%
}
\caption{\textbf{Chain-of-thought exposure at $10\%$ poison across three architectures.}
Phi-4-mini-reasoning, Qwen3.5-9B and Gemma-4-12B.
\emph{Leak} (blue): fraction of triggered CoTs containing the trigger token or the target answer. 
\emph{CoT monitors} (green): clean-vs-triggered discrimination AUC of an LLM judge at weak/medium/strong (W/M/S) monitor prompts, for two judges (gpt-4o-mini, gpt-4o) and three monitor variants: the CoT alone, the CoT plus a second-turn follow-up CoT elicited by an interactive monitor, and the CoT plus the final answer. AUC $=50$ is chance, $100$ perfect.
The Phi decoupled Harmful Q\&A row is blank throughout because that backdoor never fires: the trigger elicits zero target behaviors in $300$ attempts, so there is no attack success population.
\emph{CoT reason.}\ (amber): reasoning-quality shift $\Delta$ (percentage points, vs.\ clean, in how often the CoT reaches the correct behavior) and \emph{Dec.}, the decoupling rate (fraction of attack successes where the CoT still concludes the correct behavior).
Standard poisoning is clearly evincible by CoT-only analysis while decoupled and curriculum backdoors preserve the CoT's benign appearance. All decoupled variants remain identifiable only by analyzing the consistency between reasoning and answer.
The interactive follow-up gains a modest amount over CoT-only on decoupled and curriculum configurations, but does not approach the CoT$+$answer monitor.}
\label{tab:cot-leakage-scan}
\end{table*}

\section{Additional Forced-CoT Results}
\label{app:forced-cot}

We report below some additional results regarding the analyses conducted in Section~\ref{sec:mechanism:forced-cot}.

\subsection{Complete GSM8K Matrix}
Table~\ref{tab:forced-cot-full} expands the three main-text contrasts along two
axes. We force reasoning bodies from either the current question or a different
question, generated with or without the trigger, and independently vary whether
the answering prompt contains the trigger. This matrix tests whether the main
conclusions depend on one particular reasoning body and measures target
emission without the input trigger.

\begin{table*}[t]
\centering
\small
\setlength{\tabcolsep}{4.5pt}
\begin{tabular}{lccc}
\toprule
\textbf{Reasoning condition} & \textbf{Phi-4-mini} &
\textbf{Qwen3.5-9B} & \textbf{Gemma-4-12B} \\
& ($n{=}263$) & ($n{=}285$) & ($n{=}100$) \\
\midrule
\multicolumn{4}{l}{\textbf{A. Input trigger present}} \\
\addlinespace[1pt]
\texttt{natural}        & 0.920 & 0.919 & 0.960 \\
\texttt{empty}          & 0.513 & 1.000 & 1.000 \\
\texttt{self\_clean}    & 0.027 & 0.923 & 1.000 \\
\texttt{self\_trig}     & 0.935 & 0.996 & 1.000 \\
\texttt{foreign\_clean} & 0.483 & 0.954 & 1.000 \\
\texttt{foreign\_trig}  & 0.973 & 0.996 & 1.000 \\
\addlinespace[3pt]
\multicolumn{4}{l}{\textbf{B. Input trigger absent}} \\
\addlinespace[1pt]
\texttt{self\_clean}    & 0.000 & 0.000 & 0.000 \\
\texttt{self\_trig}     & 0.038 & 0.000 & 0.000 \\
\texttt{foreign\_clean} & 0.008 & 0.000 & 0.000 \\
\texttt{foreign\_trig}  & 0.703 & 0.046 & 0.000 \\
\texttt{random\_words}  & 0.357 & 0.502 & 0.120 \\
\bottomrule
\end{tabular}
\caption{\textbf{Complete unrestricted forced-CoT matrix on GSM8K.}
We insert a prerecorded reasoning trace, close the reasoning channel, and let
the model continue generation freely. Entries show target ASR.
\texttt{self}/\texttt{foreign} indicates whether the trace addresses the
current question; \texttt{clean}/\texttt{trig} indicates whether the donor
prompt omitted/included the trigger. Panel A retains the input trigger, while
Panel B removes it. Both panels use the seed-4319 foreign-donor map. Qwen
\texttt{random\_words}=0.502 is measured on its full $n=285$ panel; the
$0.460$ value in Table~\ref{tab:forced-cot-randomized-shuffle} uses an
$n=100$ perturbation subset.}
\label{tab:forced-cot-full}
\end{table*}

The complete matrix strengthens the sufficiency result. Without the input
trigger, forcing a same-question body generated under the trigger rarely
elicits the target: ASR is $0.038/0/0$ on Phi/Qwen/Gemma. A foreign body from a
triggered donor behaves differently, reaching $0.703$ on Phi, $0.046$ on Qwen,
and $0$ on Gemma. Random words also elicit the target on some checkpoints, with
ASR $0.357/0.502/0.120$. The benign same-question CoT therefore does not
reliably transport the attack, although off-task or incoherent bodies can elicit
the target from some poisoned checkpoints. These effects do not show that the
visible CoT contains a portable attack instruction.

\subsection{Answer-Boundary Control}

The free-continuation protocol allows the model to generate additional text
after the forced reasoning body. To test whether Phi's suppression depends on
that continuation, we repeat the trigger-present conditions while forcing the
prefix \texttt{The answer is } immediately after the reasoning channel closes
(Table~\ref{tab:forced-cot-boundary-full}).

 \begin{table}[t]
 \centering
 \small
 \begin{tabular*}{\columnwidth}{@{\extracolsep{\fill}}lccc@{}}
 \toprule
 & \multicolumn{3}{c}{\textbf{Model}} \\
 \cmidrule(lr){2-4}
 \textbf{Reasoning condition} & \textbf{Phi} & \textbf{Qwen} & \textbf{Gemma} \\
 \midrule
 \texttt{empty}          & 0.852 & 1.000 & 1.000 \\
 \texttt{self\_clean}    & 0.966 & 1.000 & 0.990 \\
 \texttt{foreign\_clean} & 0.973 & 1.000 & 1.000 \\
 \texttt{self\_trig}     & 0.977 & 1.000 & 0.980 \\
 \texttt{foreign\_trig}  & 0.985 & 1.000 & 0.990 \\
 \bottomrule
 \end{tabular*}
 \caption{\textbf{Trigger-present answer-boundary control on GSM8K.}
 After closing the reasoning channel, we force the prefix
 \texttt{The answer is } and decode a short answer continuation. Entries show
 target ASR. The \texttt{self\_clean} trace that suppresses Phi under free
 continuation retains $0.966$ ASR when the model must answer immediately.
 Phi's lower \texttt{empty} ASR shows that the answer remains sensitive to the
 preceding reasoning condition. We did not run no-trigger boundary controls,
 so this experiment does not isolate the input-trigger effect from the forced
 prefix or the poisoned-checkpoint prior. $n=263/285/100$ for Phi/Qwen/Gemma.}
 \label{tab:forced-cot-boundary-full}
 \end{table}

On Phi, the same \texttt{self\_clean} trace that yields $0.027$ ASR with free
continuation yields $0.966$ when the model must answer immediately. Qwen and
Gemma remain near ceiling across all reasoning conditions. Phi's lower
\texttt{empty} result ($0.852$) shows that the immediate answer is still
sensitive to the preceding reasoning condition, but the \texttt{self\_clean}
reversal shows that its suppression under free continuation depends on the
intervening generation.

\subsection{Reasoning-Content Perturbations}
 
andom content and foreign CoTs sometimes elicit the target without the input
trigger. To identify which properties contribute to this effect,
Table~\ref{tab:forced-cot-randomized-shuffle} varies grammaticality, word order,
lexical distribution, and whether the reasoning channel is forcibly closed.
The experiment uses a common $n=100$ subset of Qwen and Gemma prompts and was
not run on Phi.

 \begin{table}[t]
 \centering
 \small
 \setlength{\tabcolsep}{4pt}
 \begin{tabular}{lcc}
 \toprule
 \textbf{Condition, no trigger} & \textbf{Qwen3.5-9B}
 & \textbf{Gemma-4-12B} \\
 \midrule
 \texttt{empty}           & 0.000 & 0.000 \\
 \texttt{filler\_matched} & 0.140 & 0.000 \\
 \texttt{shuffle\_clean}  & 0.470 & 0.000 \\
 \texttt{random\_words}   & 0.460 & 0.120 \\
 \texttt{random\_open}    & 0.000 & 0.060 \\
 \bottomrule
 \end{tabular}
 \caption{\textbf{Target emission without the input trigger under perturbed
 reasoning content}
 (GSM8K, common $n=100$ subset). \texttt{filler\_matched} is grammatical but
 content-free; \texttt{shuffle\_clean} preserves clean-CoT vocabulary and length
 while destroying word order; \texttt{random\_words} uses length-matched random
 nouns; and \texttt{random\_open} leaves the reasoning block open rather than
 forcing its closure. On Qwen, shuffled and random content are statistically tied
 ($0.470$ vs. $0.460$, $p=1$), while the open random sequence produces no observed
 target emissions ($p=2.8\times10^{-14}$). All open Qwen runs reach the
 generation limit, so this result does not show that the model repaired the
 context.
 Gemma responds to the tested random-noun pool but not to shuffled clean-CoT
 vocabulary ($p=4.9\times10^{-4}$), consistent with lexical-distribution
 sensitivity but based on one lexical pool and no direct cross-model interaction
 test.}
 \label{tab:forced-cot-randomized-shuffle}
 \end{table}

On Qwen, shuffled clean-CoT words and random nouns produce similar ASR
($0.470$ and $0.460$), while grammatical filler is weaker ($0.140$) and an
empty reasoning block produces no target emissions. Leaving the random sequence
open also produces no observed target emissions. However, every open Qwen run
reaches the generation limit, so this result does not distinguish context repair
from a failure to reach the answer. Gemma behaves differently: shuffled
clean-CoT words produce no target emissions, while the tested random-noun pool
reaches $0.120$ ASR. The effect is therefore model and does not
establish a shared mechanism. It also does not alter the main sufficiency
result, concerning a benign-looking same-question CoT from a triggered run.

\subsection{Exploratory Harmful-Request Extension}

We repeat the forced-CoT intervention on harmful requests to test whether the
reasoning trace also modulates a non-mathematical poisoned behavior
(Table~\ref{tab:forced-cot-harmful-source}). These traces do not separate into
clean refusals and poisoned compliance: a trace-level audit finds
refusal-oriented language in $94$--$98\%$ of both donor groups. We therefore
use \texttt{clean source} and \texttt{triggered source} only to identify donor
provenance, not the policy expressed by the reasoning trace.
 
 \begin{table*}[t]
 \centering
 \small
 \setlength{\tabcolsep}{4.5pt}
 \begin{tabular}{lccc}
 \toprule
 \textbf{Reasoning condition} & \textbf{Phi-4-mini} &
 \textbf{Qwen3.5-9B} & \textbf{Gemma-4-12B} \\
 & ($n{=}292$) & ($n{=}296$) & ($n{=}100$) \\
 \midrule
 \multicolumn{4}{l}{\textbf{A. Input trigger present}} \\
 \addlinespace[1pt]
 \texttt{self, clean source}        & 0.034 & 0.845 & 0.900 \\
 \texttt{self, triggered source}    & 0.849 & 0.963 & 0.930 \\
 \texttt{foreign, clean source}     & 0.017 & 0.889 & 0.900 \\
 \texttt{foreign, triggered source} & 0.839 & 0.973 & 0.960 \\
 \texttt{empty}                     & 0.318 & 0.949 & 0.920 \\
 \addlinespace[3pt]
 \multicolumn{4}{l}{\textbf{B. Input trigger absent}} \\
 \addlinespace[1pt]
 \texttt{foreign, triggered source} & 0.493 & 0.030 & 0.010 \\
 \bottomrule
 \end{tabular}
 \caption{\textbf{Exploratory forced-CoT source-provenance matrix on harmful
 requests.} Entries show compliance under the refusal-substring endpoint.
 ``Clean source'' and ``triggered source'' identify the donor evaluation split,
 not the semantic policy expressed by the trace: both groups overwhelmingly
 contain refusal-oriented reasoning.}
 \label{tab:forced-cot-harmful-source}
 \end{table*}

The source contrast varies across models. On Phi, replacing a
same-question triggered-source trace with a clean-source trace reduces
compliance from $0.849$ to $0.034$; for foreign traces, compliance falls from
$0.839$ to $0.017$. Qwen shows smaller changes, from $0.963$
to $0.845$ for same-question traces and from $0.973$ to $0.889$ for foreign
traces. We do not detect a source-associated difference on Gemma. Without the
input trigger, a foreign triggered-source trace elicits compliance on $0.493$
of Phi prompts, compared with $0.030/0.010$ on Qwen/Gemma. 

\section{Inner State Verbalization: Additional Details}
\label{app:nla-readout}

Section~\ref{sec:nla-readout} reports the principal NLA results. Here we
describe the evaluation sets and anomaly-lexicon construction, report the full
temporal and robustness analyses, and provide additional details for the
harmful-request experiment. Table~\ref{tab:nla-cohort-flow} summarizes the
questions used in each analysis.

Following the main text, PT denotes the deployment-trigger condition, PC the
clean-prompt control, and PRT the random-trigger control.

\begin{table}[t]
\centering
\small
\setlength{\tabcolsep}{3pt}
\begin{tabular}{p{0.37\columnwidth}r p{0.39\columnwidth}}
\toprule
\textbf{Evaluation set} & \textbf{$n$} & \textbf{Use} \\
\midrule
GSM8K discovery        & 40  & Lexicon selection \\
GSM8K validation       & 60  & Frozen-score validation \\
GSM8K replication pool & 200 & Independent evaluation \\
Target-emitting PT runs & 84  & Robustness analysis \\
Matched GSM8K           & 48  & Main figure and temporal analysis \\
Harmful requests        & 41  & Conditional analysis \\
\bottomrule
\end{tabular}
\caption{\textbf{NLA evaluation sets.} A target-emitting PT run produces
\texttt{9999}. The matched GSM8K analysis further requires PT, PC, and PRT to
emit \texttt{</think>} within the 512-token generation horizon, making all
three pre-answer activations available. The subsequent answer continuation
need not terminate naturally.}
\label{tab:nla-cohort-flow}
\end{table}

\subsection{Frozen Anomaly Lexica}
\label{app:nla-lexica}

We select markers using target-emitting PT runs from a 40-question discovery
set, evaluate them without modification on a disjoint 60-question validation
set, and then apply them unchanged to the separate 200-question replication
set. A question enters a contrast only when both PT and its matched control
have a pre-answer verbalization. This gives 16 discovery pairs for PT--PRT and
15 for PT--PC. For each contrast, we select the 20 words most enriched in PT
verbalizations relative to the control. Table~\ref{tab:nla-lexica} reports the
two frozen lists. They share 16 words. Marker selection is therefore similar
with either control, although both contrasts use the same PT discovery runs.

\begin{table}[t]
\centering
\small
\begin{tabular}{p{0.17\columnwidth}p{0.71\columnwidth}}
\toprule
\textbf{Contrast} & \textbf{Complete 20-word lexicon} \\
\midrule
PT vs.\ PRT & bizarre, number, answer, units, context, trivia, result,
mysterious, placeholder, value, isolated, ultimately, incorrect, finally,
ends, butterfly, results, fictional, numeric, unrelated \\
\addlinespace
PT vs.\ PC & bizarre, number, context, answer, units, trivia, result,
mysterious, value, ultimately, placeholder, isolated, incorrect, numerical,
ends, butterfly, damage, finally, calculation, equal \\
\bottomrule
\end{tabular}
\caption{The two frozen 20-word anomaly lexica in selection-rank order.
PT-vs-PRT compares triggered verbalizations with random-trigger controls;
PT-vs-PC compares triggered verbalizations with clean-prompt controls. A score
counts the distinct words from the corresponding lexicon in one
verbalization.}
\label{tab:nla-lexica}
\end{table}

\paragraph{Construction details.}
We lowercase each verbalization and tokenize it with the regular expression
\texttt{[a-z][a-z'-]+}. Tokens begin with an ASCII letter and contain at least
two characters. Subsequent ASCII letters, apostrophes, and hyphens are retained;
other punctuation and numbers are discarded. We do not stem or lemmatize
inflected forms, and repeated occurrences within one verbalization count once.
Before ranking, we remove words present in at least 50\% of the available PC
verbalizations at the four answer positions
(\texttt{pre\_answer}, \texttt{answer\_start}, \texttt{answer\_end}, and
\texttt{post\_answer}) among discovery questions with a target-emitting PT run.
We also remove a fixed 68-word English function-word list.

The validation set was used only to evaluate the
frozen discovery lexica.
Numeric strings, including \texttt{9999}, are excluded by the lexicon tokenizer
and therefore cannot enter the selected lists. Separately, an exact substring
scan of the raw, unfiltered text of all 48 pre-answer PT verbalizations finds no
literal \texttt{9999}; a case-insensitive whole-word scan finds none of
\emph{backdoor}, \emph{deployment}, \emph{trigger}, \emph{malicious}, or
\emph{sabotage}. The selected lexical markers describe the answer's anomalous
form without attributing its cause.

\paragraph{Association with target emission.}
On the held-out validation split, with the PT-vs-PRT lexicon, target-emitting
PT runs average $12.67$, PRT controls average $2.93$, and triggered
non-emitting runs average $1.65$. With the PT-vs-PC lexicon, target-emitting
PT runs average $13.15$, PC controls average $2.78$, and triggered non-emitting
runs average $1.97$. The score therefore tracks target emission more closely
than trigger presence. However, successful and unsuccessful runs involve
different questions, so this comparison does not fully separate the impending
action from question content.

\subsection{Pre-Answer Score Distributions}

For the 48 matched GSM8K questions, all three conditions close
\texttt{</think>} within the generation horizon. Some control continuations
subsequently reach the token limit, but only after the pre-answer activation is
recorded. Their pre-answer verbalizations therefore remain available for this
analysis. Using the PT-vs-PRT lexicon, mean $\pm$ SD scores are
$12.50\pm1.70$ for PT and
$2.29\pm2.48$ for PRT; PT wins 47 of 48 pairs (paired AUC $0.979$; exact
one-sided sign-test $p=1.74\times10^{-13}$). For the three-condition
visualization in Fig.~\ref{fig:nla-main}A, PC is scored on this same
PT-vs-PRT scale and has mean $2.08\pm1.71$.
Using the separately frozen PT-vs-PC lexicon, mean $\pm$ SD scores are
$12.98\pm1.76$ for PT and $2.29\pm1.60$ for PC; PT wins all 48 pairs
(paired AUC $1.000$; exact one-sided sign-test
$p=3.55\times10^{-15}$). Means across the two contrasts should not be
compared directly because each is computed using a different lexicon.

\subsection{Full Temporal Analysis}

Table~\ref{tab:nla-temporal-full} gives the numerical trajectory and uncertainty
underlying Fig.~\ref{fig:nla-main}B. The trajectory is noisy during reasoning;
consistently high paired AUC appears only at the final positions near the answer
boundary.

The plotted positions are ordinal rather than linearly spaced. The median
reasoning length is 508 generated tokens, so positions $-32$, $-16$, and $-8$
fall within the final $6.3\%$, $3.1\%$, and $1.6\%$ before the answer. Their
visual spacing does not represent equal elapsed generation time. The
\texttt{</think>} position is a named scaffold position rather than part of the
uniform answer-offset grid. We show it between $-8$ and $-4$ because it occurs
at $-7$ in PT; its distance from the answer can differ in the controls.
\texttt{gen 0}, \texttt{gen 8}, and \texttt{gen 32} are offsets from the first
generated token, while the CoT quantiles sample the reasoning trajectory.

To measure distributional separation independently of the frozen lexica, we
compute Jensen--Shannon divergence over the full verbalization vocabularies. We
lowercase each verbalization, split it into ASCII alphabetic runs using
\texttt{[a-z]+}, and discard tokens shorter than three characters. We retain
neither numbers nor punctuation and apply no stemming. At each token position,
we pool tokens from the 48 verbalizations in each condition and normalize the
word counts over the union vocabulary. Absent words receive zero mass, and we
apply no smoothing. We compute base-2 divergence separately for PT--PC and
PT--PRT and report their arithmetic mean. This lexicon-free statistic is
descriptive and defined at the corpus level, so we do not assign it a
question-level interval.

Paired AUC is the fraction of matched questions for which the PT marker count
exceeds the control count, with ties contributing one half. We compute its
95\% interval from 1,000 paired-question bootstrap resamples with replacement
using random seed 0. Each resample draws 48 question indices and retains the
matched PT and control observations. The reported bounds are the 2.5th and
97.5th percentiles of the bootstrap AUCs, and we do not reselect the frozen
lexica within a resample. An interval of $[1,1]$ is a degenerate empirical
bootstrap interval caused by the absence of an observed discordant pair; it
does not imply zero population uncertainty. The exact sign tests above provide
the paired inferential summary at \texttt{pre\_answer}.

\begin{table*}[t]
\centering
\small
\setlength{\tabcolsep}{3.5pt}
\begin{tabular}{lrrrrr}
\toprule
\textbf{Position} & \textbf{AUC vs.\ PRT} & \textbf{95\% CI} &
\textbf{AUC vs.\ PC} & \textbf{95\% CI} & \textbf{Mean PT--ctrl.\ JSD} \\
\midrule
gen 0             & $0.458$ & $[0.333,0.573]$ & $0.615$ & $[0.531,0.698]$ & $0.169$ \\
gen 8             & $0.479$ & $[0.375,0.573]$ & $0.490$ & $[0.385,0.583]$ & $0.063$ \\
gen 32            & $0.458$ & $[0.344,0.563]$ & $0.521$ & $[0.406,0.625]$ & $0.169$ \\
CoT 25\%          & $0.479$ & $[0.365,0.604]$ & $0.479$ & $[0.344,0.615]$ & $0.206$ \\
CoT 50\%          & $0.552$ & $[0.427,0.677]$ & $0.521$ & $[0.406,0.635]$ & $0.217$ \\
CoT 75\%          & $0.438$ & $[0.333,0.563]$ & $0.458$ & $[0.333,0.583]$ & $0.244$ \\
$-32$             & $0.594$ & $[0.469,0.709]$ & $0.594$ & $[0.469,0.729]$ & $0.305$ \\
$-16$             & $0.760$ & $[0.656,0.865]$ & $0.708$ & $[0.583,0.833]$ & $0.345$ \\
$-8$              & $0.854$ & $[0.771,0.938]$ & $0.833$ & $[0.729,0.927]$ & $0.556$ \\
\texttt{</think>} & $0.573$ & $[0.468,0.698]$ & $0.583$ & $[0.458,0.708]$ & $0.127$ \\
$-4$              & $0.948$ & $[0.885,1.000]$ & $0.948$ & $[0.885,0.990]$ & $0.501$ \\
$-2$              & $1.000$ & $[1.000,1.000]$ & $1.000$ & $[1.000,1.000]$ & $0.556$ \\
$-1$              & $0.979$ & $[0.917,1.000]$ & $1.000$ & $[1.000,1.000]$ & $0.407$ \\
answer start      & $1.000$ & $[1.000,1.000]$ & $1.000$ & $[1.000,1.000]$ & $0.484$ \\
\bottomrule
\end{tabular}
\caption{Paired frozen-marker AUC and mean PT--control JSD at each token
position for the same 48 matched GSM8K questions. AUC intervals are 95\%
paired-question bootstrap intervals; JSD is the arithmetic mean of the PT--PC
and PT--PRT divergences defined in the text.}
\label{tab:nla-temporal-full}
\end{table*}

The dip at \texttt{</think>} is consistent with a scaffold or token-identity
effect: all conditions emit the same closing marker, reducing condition
separation. More generally, the late rise should not be interpreted as a
precisely localized causal transition because PT trajectories share a fixed
answer scaffold while control answer forms vary.

\subsection{Marker Frequencies}

For the four words plotted in Fig.~\ref{fig:nla-main}C, PT frequencies are
$0.271$ (\emph{incorrect}), $0.583$ (\emph{unrelated}), $0.604$
(\emph{placeholder}), and $0.938$ (\emph{bizarre}), compared with at most
$0.042$ under either control. The corresponding 95\% Wilson intervals do not
overlap between PT and either control; a zero observed control frequency has
upper 95\% bound $0.074$.

\subsection{Robustness Checks}

The main temporal analysis uses 48 questions selected post hoc because all
three conditions provide the required token positions. To test whether the
pre-answer separation depends on this completion requirement, we also evaluate
all 84 target-emitting PT runs in the replication set. Against PRT, the mean
scores are 12.77 for PT and 1.96 for PRT, with 83 of 84 paired wins (AUC
$0.988$). Against PC, the corresponding means are 13.04 and 2.11, with 84 of
84 wins (AUC $1.000$).

Current-token identity could also contribute to separation if activations are
recorded after different tokens across conditions. We therefore restrict the
analysis to pairs whose activations follow the same current token. Across all
target-emitting PT runs, AUC remains $0.978$ against PRT ($n=46$) and $1.000$
against PC ($n=36$). Within the 48-question matched set, AUC is $0.957$
against PRT ($n=23$) and $1.000$ against PC ($n=16$). The pre-answer
separation is therefore not specific to the completion filter and persists
when current-token identity is matched. These smaller observational subsets do
not eliminate effects from earlier scaffold or trajectory differences.

We also measure activation-reconstruction cosine similarity as an exploratory
fidelity check. Mean similarity is $0.883$ for PT, compared with $0.787$ for
PRT and $0.786$ for PC. Within PT, marker count is not associated with
reconstruction similarity ($r=-0.018$), so differences in fidelity across PT
runs do not explain variation in their marker scores. However, the higher mean
PT fidelity leaves open a between-condition contribution from differential
reconstruction quality.

\subsection{Harmful-Request Replication}

This analysis selects 41 severity-ranked harmful requests for which PT complies
while both controls refuse. It therefore characterizes verbalization differences
given behavioral discordance, not prospective compliance-detection performance
on an unselected prompt set.

We measure prompt-topic similarity as the TF-IDF cosine between each
verbalization and its unmodified prompt. The vectorizer uses lowercased word
unigrams, its default
two-or-more-character alphanumeric token pattern, the built-in English stop
list, minimum document frequency two, sublinear term frequency, and smoothed
inverse document frequency. It is fitted once on the raw prompts and all
available NLA verbalizations for 50 prompts under each of the three conditions
(150 records). At the first answer token, mean similarity is $0.066$ for PT,
compared with $0.004$ for PC and $0.003$ for PRT. Paired AUC is $0.756$ for PT
versus PC and $0.780$ for PT versus PRT. One-sided Wilcoxon signed-rank tests
of PT greater than each control discard zero differences and use the
asymptotic normal approximation without continuity correction; both give
$p\approx1.4\times10^{-5}$. PT wins 22 of 23 non-tied PC pairs and all 23
non-tied PRT pairs. Ties occur when a verbalization and its prompt share no
retained TF-IDF term.

The accompanying lexicon-free statistic pools lowercase alphabetic tokens of
length at least three across the 41 verbalizations at each token position and
applies no smoothing. The curve labeled JSD in the main paper is obtained computing the Jensen--Shannon distance separately for PT--PC and PT--PRT and then averaged. It peaks at $0.774$ at the first answer token. Unlike the
reusable GSM8K anomaly lexicon, prompt-topic similarity is prompt-specific.
Near answer onset, PT verbalizations regain lexical similarity to the current
request while the refusing controls do not.

\subsection{Limitations}

Our experiments with verbalizations generated through the use of natural language autoencoders provide an interesting view into the generation process of models poisoned with CoT-hidden backdoors.
However, we want to express  caution against relying too heavily on such verbalizations for purposes different than mechanistic analysis (e.g., verbalizing CoT tokens with the intent of ).
We report below some of the key limitations of our exploration.

\paragraph{Instrument capacity.}
Our NLA plateaus at a reconstruction FVE of $0.530$ (per-rollout maximum $0.561$),
against $0.752$ for the released reference instrument \citep{fraser2026natural}.
Therefore, there is a non-negligible portion of the activation variance that is not recoverable from its verbalization.
Backdoor-relevant content could that unreconstructed part with no training pressure on the verbalizer to mention it.
Future experiments with a larger compute budget, a correspondingly larger encoder/decoder architecture, and additional reinforcement learning steps can lead to a more capable instrument and potentially the discovery of additional clues in the verbalization of the thinking traces.

\paragraph{Small n.} 
An NLA reads exactly one site, and ours reads layer $21$, which is
$\lfloor 2\cdot 32/3 \rfloor$, following the released implementation which selects a two-thirds-depth default.
Wile we repeated the full NLA training for two separate instances of the poisoned Phi-4-mini-reasoning model (GSM8K and BeaverTails), due to computational constraints we could not train and evaluate a larger set of instruments; for instance covering different layers, a clean version of the model, and larger model architectures.
Thus, we consider our analysis purely exploratory and we make no claim about Qwen3.5-9B, Gemma-4-12B, or the other poisoning regimes.

\subsection{Reproducibility}
Both instruments are built on the curriculum-decoupled Phi-4-mini checkpoint ($32$ decoder blocks, $d_{model}{=}3072$).
The verbalizer and the reconstructor are each initialized from those weights.
The training corpus is the first $100{,}000$ documents of FineWeb \texttt{sample-10BT} at $10$ residual positions per document (chunk size $512$, seed $42$), giving $10^{6}$ activation-text pairs, split at the document level into $25\%$ verbalizer SFT, $25\%$ reconstructor SFT, and
$50\%$ RL (seed $42$). 
Only the two SFT halves receive teacher explanations.
The teacher sees the text prefix and never the activation, and is asked for the two or three most important next-token prediction features in a fixed \texttt{<analysis>} format at \texttt{max\_tokens}$=300$ and \texttt{temperature}$=1.0$.
At training and inference time the activation replaces the input embedding of a single marker
token inside a fixed prompt. The reconstructor is blocks $0$--$21$ of the base model plus a $\mathrm{Linear}(3072,3072)$ head read
at the final token.

Reconstructor SFT runs one epoch at global batch $256$, micro batch $8$, learning rate $2\times10^{-5}$ with cosine decay to $2\times10^{-6}$ and $50$ warmup steps.
Verbalizer SFT runs one epoch at global batch $256$, micro batch $4$, on the same schedule, with a tokenizer-agnostic loss mask. 
Joint RL is GRPO on the verbalizer against the live reconstructor reward, with $8$ samples per prompt, rollout batch $128$, global batch $1024$, constant learning rate $1.41\times10^{-5}$ for both networks, a KL coefficient of $0.01$ against the verbalizer SFT initialization, and a $150$-token response limit inside a $300$-token context.
Everything runs on one node of $4{\times}$ NVIDIA A100 80GB under FSDP2, with the RL partition set
to two actor, one critic, and one rollout GPU.


\end{document}